\documentclass[referee]{raa}           
\usepackage{graphicx,times}
\usepackage{natbib}
\usepackage{amssymb,amsmath}
\bibpunct{(}{)}{;}{a}{}{,}
\graphicspath{{plots/}}
\newcommand{\msun}{M_\odot}
\usepackage{hyperref}
\hypersetup{pdftitle = The title of my PDF, pdfauthor = My name, pdfsubject= The subject, pdfkeywords = keyword1 keyword2 keyword3} 
\hypersetup{colorlinks = true, linkcolor = green, anchorcolor = red, citecolor = blue, filecolor = red, pagecolor = red, urlcolor = red}

\begin{document}

   \title{Investigating the UV-excess in star clusters with $N$-body simulations: predictions for future CSST observations 
$^*$
\footnotetext{\small $*$ 
Supported by the research grants from the China Manned Space Project with No. CMS-CSST-2021-A08. ORCID ID for authors: Xiaoying Pang (0000-0003-3389-2263); Long Wang (0000-0001-8713-0366); M.B.N. Kouwenhoven (0000-0002-1805-0570).}
}

 \volnopage{ {\bf 2022} Vol.\ {\bf --} No. {\bf --}, 000--000}
   \setcounter{page}{1}

   \author{Xiaoying Pang\inst{1,2}, Qi Shu\inst{1}, Long Wang\inst{3,4,5}, and M.B.N. Kouwenhoven\inst{1}}

\institute{Department of Physics, Xi'an Jiaotong-Liverpool University, 111 Ren'ai Road, Suzhou Dushu Lake Science and Education Innovation District, Suzhou Industrial Park, Suzhou 215123, P.R. China;  {\it Xiaoying.Pang@xjtlu.edu.cn}\\
\and 
Shanghai Key Laboratory for Astrophysics, Shanghai Normal University, 100 Guilin Road, Shanghai 200234, P.R. China\\
\and 
School of Physics and Astronomy, Sun Yat-sen University, Daxue Road, Zhuhai, 519082, P.R. China \\
\and
Department of Astronomy, School of Science, The University of Tokyo, 7-3-1 Hongo, Bunkyo-ku, Tokyo 113-0033, Japan\\
\and 
RIKEN Center for Computational Science, 7-1-26 Minatojima-minami-machi, Chuo-ku, Kobe, Hyogo 650-0047, Japan\\
\vs \no
   {\small Accepted for publication at Research in Astronomy and Astrophysics on 30 July}
}

\abstract{ 
We study the origin of the UV-excess in star clusters by performing $N$-body simulations of six clusters with $N=10$k and $N=100$k (single stars \& binary systems) and metallicities of $Z=0.01$, $0.001$, and $0.0001$, using  \textsc{petar}.   All models initially have a 50 percent primordial binary fraction. 
Using \texttt{GalevNB} we convert the simulated data into synthetic spectra and photometry for the China Space Station Telescope (CSST) and Hubble Space Telescope (HST). From the spectral energy distributions we identify three stellar populations that contribute to the UV-excess: (1) second asymptotic giant branch stars, which contribute to the UV flux at early times; (2) naked helium stars, and (3) white dwarfs, which are long-term contributors to the FUV spectra. Binary stars consisting of a white dwarf and a main-sequence star are cataclysmic variable (CV) candidates. The magnitude distribution of CV candidates is bimodal up to 2\,Gyr. The bright CV population is particularly bright in $FUV-NUV$. The $FUV-NUV$ color of our model clusters is 1--2\,mag redder than the UV-excess globular clusters in M\,87 and in the Milky Way. This discrepancy may be induced by helium enrichment in observed clusters. Our simulations are based on simple stellar evolution; we do not include the effects of variations in helium and light elements or multiple stellar populations. A positive radial color gradient is present in CSST $NUV-y$ for main-sequence stars of all models with a color difference of 0.2--0.5\,mag, up to 4 half-mass radii. The CSST $NUV-g$ color correlates strongly with HST $FUV-NUV$ for $NUV-g>1$\,mag, with the linear relation $FUV-NUV=(1.09\pm0.12)\times(NUV-g)+(-1.01\pm0.22)$. This allows for conversion of future CSST $NUV-g$ colors into HST $FUV-NUV$ colors, which are sensitive to UV-excess features. We find that CSST will be able to detect UV-excess in galactic/extra-galactic star clusters with ages $>200$\,Myr. 
}
   \authorrunning{Pang et al. }            
   \titlerunning{UV-excess in star clusters}  
   \maketitle

%
\section{Introduction}           
\label{sec:intro}

Excess of UV emission between the Lyman break (1216\,$\rm \AA$) and 2500\,$\rm \AA$ in the spectral energy distribution (SED) was unexpectedly found in many early-type galaxies. This phenomenon is called the ``UV-upturn'' or the ``UV-excess'' \citep{code1979,burstein1987,oconnell1992,dorman1995}, after the first UV-upturn discovery made in the bulge of M\,31 \citep{code1979}. The strength of the UV-upturn tends to rise with increasing stellar mass of a galaxy \citep{dantas2019}, with a higher metallicity \citep{burstein1988}, and with increasing age \citep{smith2012}. The evolution of the UV-upturn with redshift is still under debate \citep{ali2018,dantas2019}.

The corresponding effective temperature for the UV-upturn feature should be above 20,000\,K \citep{brown1997}. However, most of the early-type galaxies are quiescent in their star formation \citep{welch1982,oconnell1986,buzzoni2012}. The hot O-B stars (30,000--40,000\,K) cannot contribute to the UV-excess emission \citep{buzzoni2012}. Numerous studies have been carried out to identify the candidates that contribute to the observed UV-excess. Old, hot and low-mass stars have been a popular choice. These include, for example, post-asymptotic giant branch (AGB) stars \citep{greggio1990} and regular AGB stars \citep{guerrero2020}; extreme horizontal branch stars \citep[EHB][]{yi1997,yoon2004,ree2007}; 
young white dwarfs that are hot and blue \citep{mestel1952, rauch2014,torres2014,pang2016}. On the other hand, interacting binary systems containing sub-dwarfs have also been a proposed sources of  the UV-upturn \citep{zhang2005,han2007,han2010}. 

All these candidate populations of stars are present in clustered stellar environments. UV-excess emission has been found in several old Galactic globular clusters (GCs): 47\,Tuc \citep{oconnell1997}, NGC\,6388 and NGC\,6441 \citep{rey2007}; and in the old open cluster NGC\,6791 \citep{buzzoni2012}. EHB stars are present in each of these clusters, and are thought to produce the observed UV-excess \citep{yi1999, yoon2006}. \citet{fan2020} report five star clusters with ages of around 1\,Gyr in M\,33 with UV excess. A large number of GCs with strong UV emission were revealed in M\,87 \citep{sohn2006}. The population of GCs in M\,87 is distinct from those of the Milky Way \citep{dorman1995} or Local Group counterparts. They are generally more metal-rich (near solar abundance) than Milky Way GCs, and are typically 1~magnitude bluer than the early-type galaxies in $FUV-V$ color. The EHB stars generated by the metal-rich and helium enhancement model \citep{lee2005,chung2011, bekki2012} appear to be promising candidates for the UV features found in the GCs of M\,87.

 The presence of multiple population can also dramatically affect the UV color of star clusters \citep{milone_hubble_2018,milone_multiple_2020,zennaro_four_2019,jang_integrated_2021}. A second population stars enhanced in helium from Y=0.25--0.26 up to Y$>$0.40 can produce hot and blue HB stars that are major contributors to the UV flux. These are proposed in theoretical studies and are also confirmed by spectroscopic measurements \citep{DAntona2005,tailo2019,tailo2020,mario2014,milone_hubble_2018}.  
\citet{jang_integrated_2021} has provided direct evidence of the effect of multiple populations on the integrated colors of Galactic GCs based on HST photometry. The second generation of stars with helium enrichment are bluer in the UV and are hotter than the first generation stars. However, in this work, we focus on simple stellar populations with a pristine helium and light element abundance, with the motivation to search for the stellar candidates making contribution to the UV colors of the star clusters. The inclusion of multiple stellar populations and helium variations are beyond the scope of the current study, and deserve further future investigation.

Studying the UV-excess candidate stars in a cluster environment can greatly benefit from $N$-body simulations using sophisticated codes such as those in the \textsc{NBODY6(++)} family \citep{spurzem1999,aarseth2003,spurzem2008,Nitadori2012,wang2015}, \textsc{petar} \citep{Wang2020b} or \textsc{AMUSE} \citep{PZ2013}. Such $N$-body simulations can be used to trace the dynamical history of individual stars and binary systems with high precision, can resolve the binary orbits, and can accurately model the processes of stellar evolution (wind mass loss, evolution of stellar radii and core radii) of all stars and multiple systems in star clusters.

$N$-body simulations produce physical and kinematical properties of the stellar population over time. In order to be able to compare these data with observations, \citet{pang2016} developed the code \texttt{GalevNB}, which converts fundamental physical stellar properties, such as stellar mass, temperature, luminosity and metallicity into observables for a variety of filter bands used by mainstream instruments/telescopes, such as HST, ESO, SDSS, and 2MASS, and into spectra that span from the far-UV (90 $\rm \AA$) to the near-IR (160 $\rm \mu$m). 
Combination \texttt{GalevNB} with $N$-body simulations allows for a direct comparisons between observational data and numerical results. It is also possible to trace the photometric and dynamical evolution of the individual stars or binary systems that are the candidates for the UV-upturn.  


The future Chinese Space Station Telescope (CSST) features a collecting surface with a diameter of 2~meters, observes at wavelengths ranging from the NUV at 0.255\,$\rm \mu$m to the infrared at 1.7\,$\rm \mu$m. CSST is expected to have a spatial resolution  $\lesssim 0.15''$ \citep{cao2018, gong2019}. The limiting magnitudes in the $g$-band and $NUV$-band are 26.3\,mag and 25.4\,mag respectively, which will extend down to 27.5\,mag and 26.7\,mag for the ultra-deep-field observations. CSST will thus be an essential tool for observing extragalactic objects at a later time than the Hubble Space Telescope (HST), especially for star clusters in the Local Group galaxies. 

In this study, we investigate the sources of UV-excess in star clusters using $N$-body simulations, and through employing \texttt{GalevNB} to 
obtain synthetic CSST magnitudes of six filters $NUV$ (2500-3000$\rm \AA$), $g$, $r$, $i$, $z$, $y$; and HST/Space Telescope Imaging Spectrograph (HST/STIS) $FUV$ and $NUV$ (1500-3000$\rm \AA$) filters. This enables us to analyze the spectral and photometric features of the simulated star clusters and search for the sources of UV-excess in the wavelength range of 1216--2500\,$\rm \AA$. By means of synthesizing observations from $N$-body simulations, we aim to determine the sensitivity and efficacy of the CSST for detecting UV-excess in star clusters.

This paper is organized as follows. Section~\ref{sec:nbody} introduces the $N$-body code \textsc{petar} and the initial conditions for the stellar and binary populations in our models. Section~\ref{sec:SED} presents the SEDs of the modeled star clusters and identifies the candidate stellar populations can produce the UV-excess. In Section~\ref{sec:phot}, we show the photometric features in the CSST and HST filter bands for the simulated clusters after conversion with \texttt{GalevNB}. Section~\ref{sec:binary_evol} investigates the properties of normal detached binary stars and the UV-excess candidate binary stars. In Section~\ref{sec:comp}, we compare the HST/STIS colors of the simulated clusters to those of star clusters with observed UV-excess. In Section~\ref{sec:predict} we provide predictions of the sensitivity of CSST for studying star clusters with UV-excess. Finally, we summarize our findings in Section~\ref{sec:summ}.



\section{$N$-body simulations} \label{sec:nbody}

\subsection{$N$-body code \textsc{\textsc{petar}}} \label{sec:petar}

In order to model the evolution of star clusters and their stellar populations, it is necessary to accurately follow the dynamical and stellar evolution of both stars and binary systems.
To achieve this, we use the high-performance $N$-body code, \textsc{petar} \citep{Wang2020a} to carry out star cluster simulations. \textsc{petar} combines the particle-tree and particle-particle algorithm \citep{oshino2011} and slow-down algorithmic regularization \citep{Wang2020b}. 
The former is a hybrid method that uses the Barnes-Hut tree method \citep{Barnes1986} for efficiently calculating the long-range interaction between particles (stars) and the 4th order Hermite integrator \citep[e.g.][]{aarseth2003} for accurately evaluating the short-range interactions between the constituents of the system. 
The latter is specifically designed for integrating the orbits of particles in multiple systems, including close encounters between stars, binary systems, and higher-order few-body systems. 
\textsc{petar} has been developed based on the framework for developing parallel particle simulation codes (\textsc{fdps}), which can achieve a high performance by using multi-process parallel computing \citep{Iwasawa2016,iwasawa2020,namekata2018}.

The recently-updated single and binary stellar evolution codes, \textsc{sse} and \textsc{bse}, \citep{tout1997,hurley2000,hurley2002, Banerjee2020} are both used in \textsc{petar}. 
Unlike computationally-expensive hydro-dynamical modelling of stellar evolution, these population synthesis codes provide a reasonable approximation of the stellar evolution while the computational cost is negligible when compared to that of the integration of the $N$-body system. 
The stellar evolution model of \cite{Banerjee2020} uses semi-empirical stellar wind prescriptions from \cite{Belczynski2010}. We adopt the ``rapid'' supernova model and material fallback from \cite{fryer2012}, along with the pulsation pair-instability supernova model \citep[][]{Belczynski2016} for the formation of compact objects.

\subsection{Initial conditions}\label{sec:initial}

We simulate the evolution of six star clusters using \textsc{petar}. Three models are initialized with $N=10,000$ particles (10k), and the other three with $N=100,000$ (100k) particles. The \textsc{petar} code represents each particle as an individual star in the cluster. The initial parameters for the models are set up using the  \textsc{Mcluster} version provided by \citet{kamlah2021}, which is updated of the original \textsc{Mcluster} of \citet{kupper2011}. 
 We carry out three simulations of models with $N=10$k and $N=100$k by assigning three metallicities, $Z=0.01$ (-2Z), $0.001$ (-3Z), and $0.0001$ (-4Z), representing  metal-rich, intermediate, and metal-poor star clusters. 
We refer to these six models as 10k-2Z, 10k-3Z, 10k-4Z, 100k-2Z, 100k-3Z and 100k-4Z, respectively. 


\textsc{petar} does not include gas dynamics. We therefore start simulations from virial equilibrium, after all gas has been expelled by stellar radiation and wind feedback from massive stars.
We adopt a tidal field corresponding to that of a circular orbit in the solar neighbourhood. The 10k and 100k models are evolved until 2\,Gyr and 10\,Gyr, respectively. The simulation time is shorter in lower-mass clusters, as these disrupt faster than the higher-mass star clusters. 

Models with identical particle numbers (i.e., 10k or 100k) are initialized with the same distributions of masses, positions and velocities. All cluster models follow a \cite{king1966} initial density profile with a dimensionless King parameter of $W_0=6$, and are assigned with initial half-mass radii of $R_{h,0}=2.0$~pc for the 10k models and $R_{h,0}=4.0$~pc for the 100k models. We adopt the initial mass function (IMF) of \cite{kroupa2001} for all models, with a mass range $0.08-80\,M_\odot$ for the 10k models and $0.08-100\,M_\odot$ for the 100k models.
All clusters are initialized in virial equilibrium, $Q=0.5$, where the virial ratio $Q=|T/U|$ is the ratio between the total kinetic energy ($T$) and the total potential energy ($U$) of the star cluster. We summarize the initial conditions of all models in Table~\ref{table:initail}.

\begin{table}
	\centering
	\caption{
	Initial conditions for the six star cluster models. $Z$ is the metallicity of the cluster;
	$R_{h,0}$ is the initial half-mass radius; 
	$R_{t,0}$ is the initial tidal radius. References: 1: \citet{king1966}; 2: \citet{kroupa2001}.}
	\label{table:initail}	
	\begin{tabular}{lllllll} 
		\hline
		  Model ID &  10k-2Z & 10k-3Z & 10k-4Z  &  100k-2Z & 100k-3Z & 100k-4Z \\
		\hline
		        Profile$^1$ & 
		         $W_0$ = 6 &
		         $W_0$ = 6 & 
		         $W_0$ = 6 &
		         $W_0$ = 6 &
		         $W_0$ = 6 &
		         $W_0$ = 6 \\
		\hline
				IMF$^2$  & 
				 0.08 - 80 $\rm \msun$ & 
				 0.08 - 80 $\rm \msun$ & 
				 0.08 - 80 $\rm \msun$ & 
				0.08 - 100 $\rm \msun$ & 
				0.08 - 100 $\rm \msun$ & 
				0.08 - 100 $\rm \msun$ \\ 
		\hline
				$M$ ($\rm \msun$) &
				9\,069 & 9\,069 & 9\,069 &
				89\,853 & 89\,853 & 89\,853\\
		\hline
				$Z$ &
				0.01 & 0.001 & 0.0001 & 0.01 & 0.001 & 0.0001 \\ 
		\hline
				$R_{h,0}$ (pc) & 
				2.0 & 2.0 & 2.0 & 4.0 & 4.0 & 4.0 \\
		\hline
				$R_{t,0}$ (pc) &
				23.5 & 23.5 & 23.5 & 49.9 & 49.9 & 49.9 \\
		\hline
				Time (Gyr) &
				2.0 & 2.0 & 2.0 & 10.0 & 10.0 & 10.0 \\
		\hline
	\end{tabular}
\end{table}

\subsection{Binary setup}\label{sec:int_binary}

All star cluster models are initialized with a 50\% primordial binary fraction {\bf ($B/(B+S)$)}, i.e., the total number of single stars ($S$) in each cluster equals the total number of binary systems ($B$) in the cluster. For example, in the $N=S+B=10$k models, there are $S=5,000$ single stars and $B=5,000$ primordial binary systems (i.e., a total of $2B=10,000$ stars in binary systems). The primordial binary systems are assigned a mass ratio, a semi-major axis, and an eccentricity as described below. We define the mass ratio of a binary system consisting of two stars of masses $m_1$ and $m_2$ as $q=m_2/m_1$, where $m_2\leq m_1$. 

For stellar masses below 5\,$\rm M_\odot$, 
the masses of both binary components are generated through random pairing from the IMF.
For stars with masses above 5\,$\rm M_\odot$, the mass of the primary star ($m_1$) is randomly selected from the IMF, and the secondary mass $m_2=q\,m_1$ is subsequently assigned after drawing the mass ratio from a uniform distribution \citep[$0.1<q<1$,][]{sana2012} in the mass range $0.08 \,M_\odot \leq m_2 \leq 80 (100)\,M_\odot$.
As a consequence of this initial setup, the number of massive stars ($>5\,M_\odot$) in binary systems is roughly double that of the number of massive single stars. We refer to \cite{kouwenhoven2009} for an extensive discussion on how the choice of random pairing affects the binary fraction and the mass ratio distribution for different primary star mass ranges.

The initial semi-major axis distribution of the primordial binary systems follows a uniform distribution in $\log a$ in the range $0.05-50$~AU. The lower limit of the semi-major axis distribution arises from the physical size of the stars, while the upper limit is comparable to the hard-soft boundary in star clusters  based on Heggie-Hills law \citep{heggie1975,Hills1975}. The initial eccentricity distribution is thermal: $f(e) \propto e$ \citep[see, e.g.,][]{heggie1975,goodman1993}. We adopt the eigenevolution and feeding algorithms of \citet{kroupa2013}. 
All binary systems are assigned random spatial orientations and a random orbital phase.


\section{Spectra Energy Distributions} \label{sec:SED}

After having completed the simulations of the six models, we use \texttt{GalevNB} \citep{pang2016} to convert the theoretical data produced by the $N$-body simulations (stellar mass, temperature, luminosity, and metallicity) into observational magnitudes of HST/STIS and CSST filters, and into spectra that span from far-UV to near-IR wavelengths. Individual spectra are produced for each star by selecting or interpolating template spectra from the \citet{lejeune1997} spectral library. \texttt{GalevNB} sums up the fluxes of individual stars and produces the integrated SEDs of the simulated clusters. In order to quantify the individual contributions of particular stellar populations to the integrated SED of a cluster, we also separately sum up the SEDs for specific populations, such as single stars, binary systems, and stars of a certain spectral types. The abbreviations for the different spectral types are listed  in Table~\ref{table:SP}. 

\begin{table}[h]
\begin{center}
\caption[]{Stellar types defined in \textsc{petar}}\label{Table 3}
\label{table:SP}
 \begin{tabular}{cc}
  \hline\noalign{\smallskip}
 Integer
  &  Stellar type                \\
  \hline\noalign{\smallskip}
0  & Deeply or fully convective low-mass Main sequence star ($\le0.7\,M\odot$) (LMS)  \\ 
1  & Main sequence star with mass $\ge0.7\,M\odot$ (MS)           \\
2  & Hertzsprung Gap (HG)          \\
3  & First Giant Branch (GB) \\
4  & Core Helium Burning (CHeB) \\
5  & First Asymptotic Giant Branch (FAGB) \\
6  & Second Asymptotic Giant Branch (SAGB) \\
7  & Main Sequence Naked Helium star (HeMS) \\
8  & Hertzsprung Gap Naked Helium star (HeHG) \\
9  & Giant Branch Naked Helium star (HeGB) \\
10 & Helium white dwarf (HeWD) \\
11 & Carbon/Oxygen White Dwarf (COWD) \\
12 & Oxygen/Neon White Dwarf (ONWD) \\
13 & Neutron Star (NS) \\
14 & Black Hole (BH)\\
15 & Massless Supernova (SN) \\
  \noalign{\smallskip}\hline
\end{tabular}
\end{center}
\end{table}

\subsection{Candidate stars contributing to the UV-excess}\label{sec:candidates}

\begin{figure*}[tb!]
\centering
\includegraphics[angle=0, width=1.\textwidth]{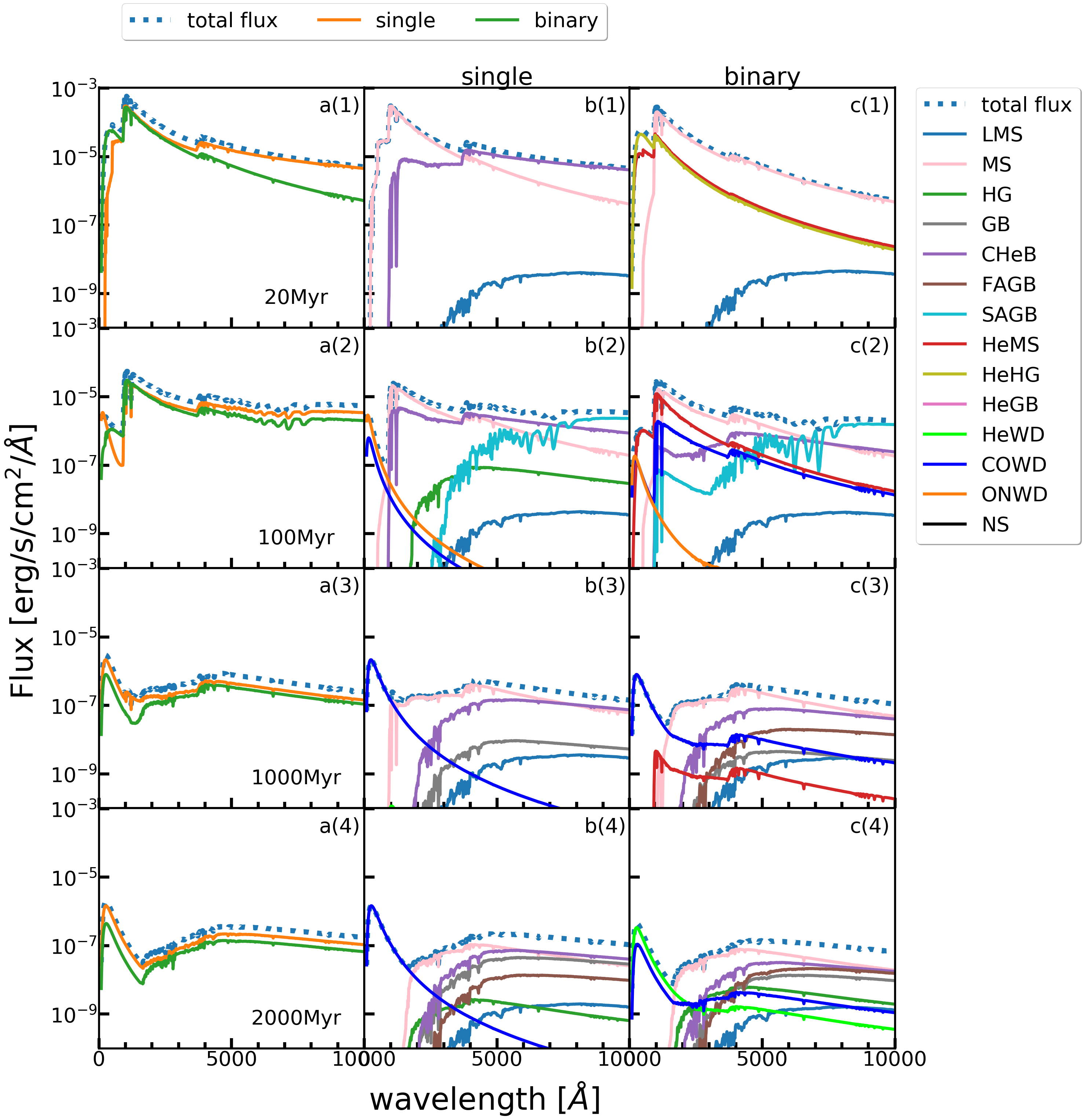}
    \caption{The spectral energy distribution (SED) of model 10k-2Z. Panels~(a) are for the entire cluster (dotted blue curves), for the single stars (solid orange curves), and for binary systems (solid green curves). Panels~(b) and~(c) show the SEDs for single stars and binary systems, respectively, in which we specify the contributions of populations of different spectral types (see Table~\ref{table:SP}) to the SED.}
\label{fig:sed_10k-2z}
\end{figure*}

\begin{figure*}[tb!]
\centering
\includegraphics[angle=0, width=1.\textwidth]{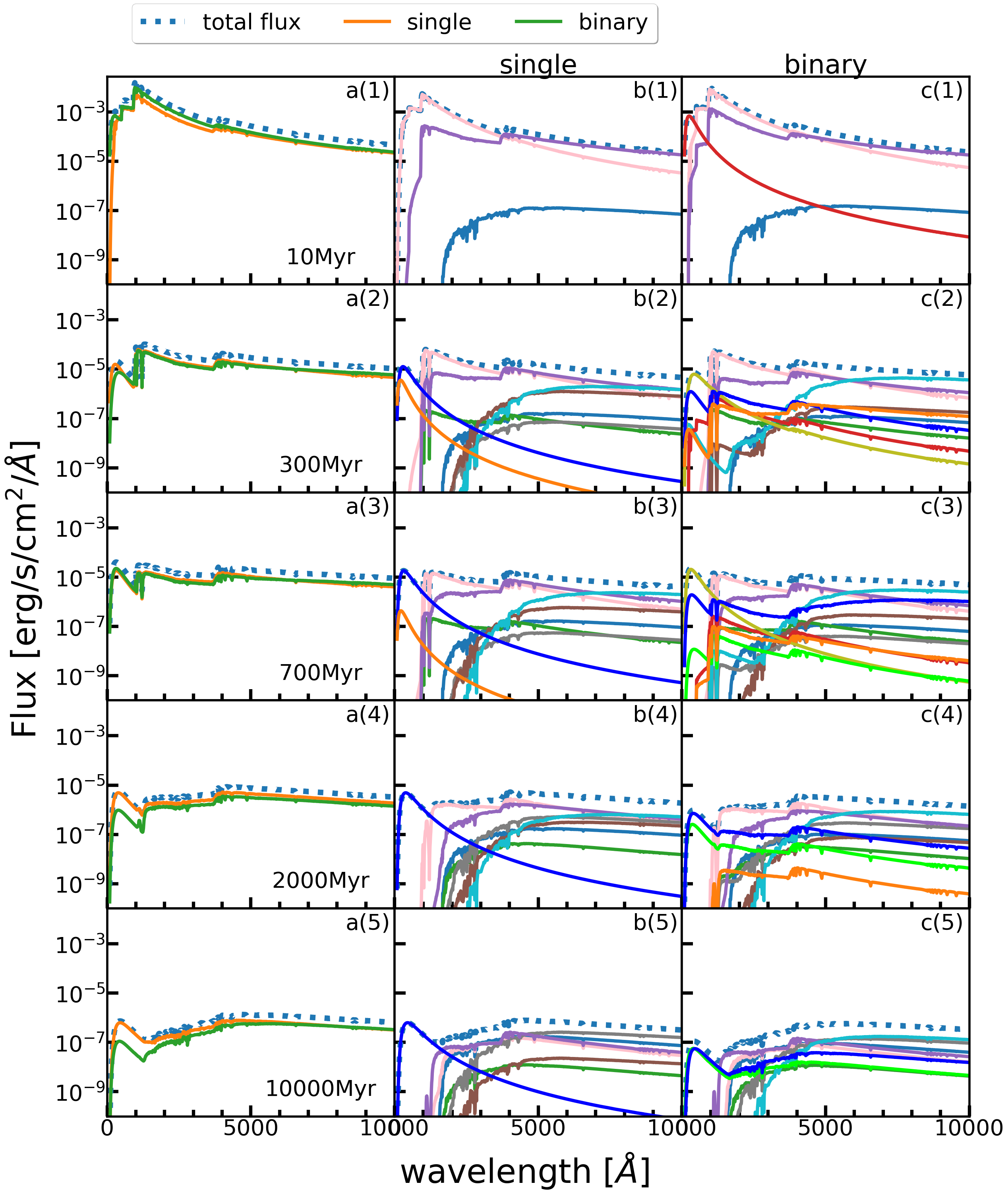}
    \caption{The SEDs of model 100k-4Z. Colors and symbols are identical to those in Figure~\ref{fig:sed_10k-2z}.}
\label{fig:sed_100k-4z}
\end{figure*}

In panels~(a) of Figures~\ref{fig:sed_10k-2z} and~\ref{fig:sed_100k-4z}, we show the SEDs of models 10k-2Z and 100k-4Z, for the entire timespan of the simulations ((a).1--(a).4). The SEDs of the other models are shown in Appendix~\ref{sec:apx_sed}. The number of individual stars in binary systems is initially double the number of single stars (see Section~\ref{sec:int_binary}). Therefore, the population of massive stars in binary systems contribute more UV flux (1216\AA$<\lambda<$2500\AA) during the first 100\,Myr. In the 10k metal-poor models (10k-3Z and 10k-4Z), binaries dominate the UV flux when $t<200$\,Myr. For the 100k models, the contribution of the binary systems to the UV-excess occurs much earlier, mainly at times $t<70$\,Myr (100k-2Z), $t<80$\,Myr (100k-3Z) and $t<100$\,Myr (100k-4Z), respectively.

\subsubsection{Second Asymptotic Giant Branch (SAGB)} \label{sec:SAGB}

In order to identify the UV-excess candidate stars, we plot the SEDs of single stars (panels~(b) of Figures~\ref{fig:sed_10k-2z} and~\ref{fig:sed_100k-4z}) and binary systems (panels (c)), in which the contribution to the SEDs by each spectral type is indicated with  a different color. 

We identify three populations of candidate stars that provide a significant contribution to the UV-excess in the clusters. When the clusters are young, the second asymptotic giant branch (SAGB, cyan curve) stars generates a peak in the UV range when $t<200$\,Myr in the 10k models. This population continues to contribute to the UV flux until $t\approx 700$\,Myr in the 100k models. However, although their UV luminosity is high, SAGB stars are short-lived. Only when stars are sufficiently massive (4--8\,$\rm M_\odot$) they can enter the SAGB phase. At this time, the hydrogen shell is re-ignited due to a second dredge-up, and it grows together with the helium-burning shell. A helium shell flash occurs, which releases a large amount of energy \citep{hurley2000} for SAGB stars; this generates a UV peak. 

\subsubsection{Naked Helium stars (HeMS)} \label{sec:HeMS}

Another group of candidates are the Main Sequence Naked Helium star (HeMS). When massive stars ($M>15\,{\rm M_\odot}$) evolve off the MS, their helium core can ignite degenerately in a helium flash. Such stars then lose a significant amount of the mass due to strong stellar winds. Sometimes they even lose their entire outer envelopes during the phase of helium fusion. When this occurs, the helium-burning cores of such stars will be revealed before they become WDs or neutron stars. Such exposed helium-burning cores are known as HeMS \citep{hurley2000}, and can have temperatures reaching $>50,000$\,K. Due to a lack of template spectra for hot stars, \texttt{GalevNB} adopts blackbody curves to approximate the SEDs of objects with temperatures above 50,000\,K. 
The evolutionary path of a HeMS star depends on whether or not it is part of a binary system. 

In all of our simulated clusters, HeMS stars sparkle mostly in the SED of binary systems during the first 50\,Myr, and become the major contributors to the UV flux at young ages. They radiate in the UV until the cluster reaches an age of a few hundred million years. HeMS stars that are part of a binary system continue to emit UV flux until $t\approx 1$\,Gyr, although their contribution to the UV-excess declines as the cluster ages. It should be noted, however, that the number of single HeMS stars in our models is much smaller than the number of HeMS binaries. This difference originates from the initial binary setup for our modeled clusters, for which massive stars preferentially reside in binary systems.

\subsubsection{White dwarfs (WDs)} \label{sec:WD}

White dwarfs (WDs) are long-term contributors to the UV-excess (orange, blue and light-green curves in panels~(c) of Figures~\ref{fig:sed_10k-2z} and~\ref{fig:sed_100k-4z}). WDs are formed after the first 30-80\,Myr in all models. WDs appear earlier in the metal-rich models, at 30--50\,Myr, and appear at 80\,Myr in the metal-poor models. Stars with masses of 8--10.5\,$\rm M_\odot$ will end up as Oxygen-Neon WDs (ONWDs, orange curve). Stars of intermediate mass will evolve into Carbon-Oxygen WDs (COWDs, blue curve). Low-mass stars are unable to fuse helium; such stars will evolve into helium WDs (HeWD, light green curve), and appear only after the cluster is few hundred Myr old. 

Young WDs are hot and blue \citep{mestel1952,torres2014}, and radiate a substantial fraction of their energy in the UV ($<2000$\,\AA). COWDs dominate in the UV flux in the long run, while the ONWDs mainly contribute to the UV flux at younger ages. As the cluster ages beyond 1\,Gyr, HeWDs become a major contributor to the UV flux. Although WDs eventually cool down, they are continuously being formed throughout the lifetime of the clusters. Therefore, the contribution of the WDs to the UV-excess for the star cluster is sustained over long timespans, and cannot be neglected.

\subsection{UV flux ratio} \label{sec:UV_ratio}

\begin{figure*}[tb!]
\centering
\includegraphics[angle=0, width=1.\textwidth]{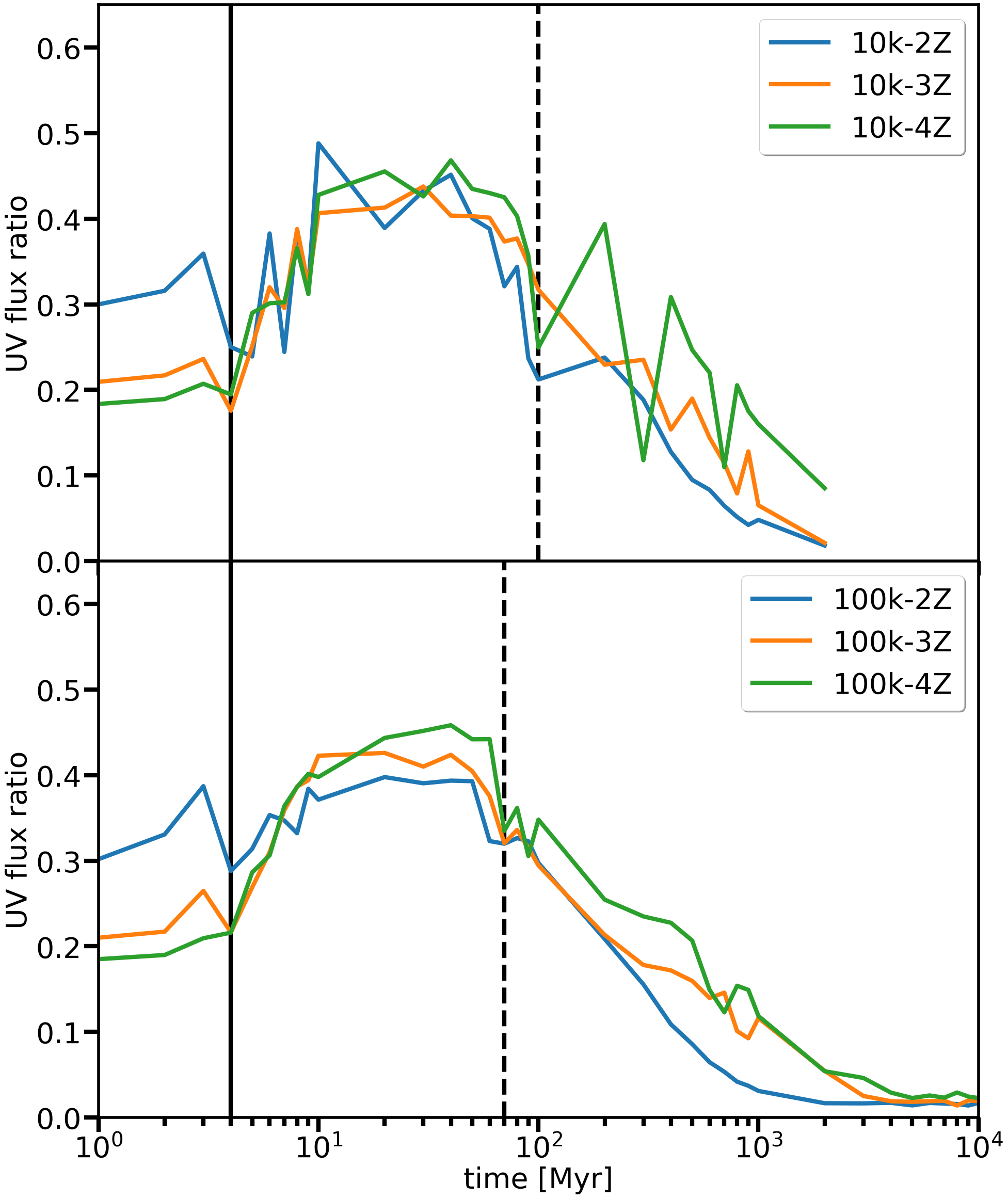}
    \caption{The evolution of the UV flux (from 1216\,$\rm \AA$ to 2500\,$\rm \AA$) ratio with time. The upper and lower panels show the results for the 10k and 100k models, respectively. 
   The vertical solid (4\,Myr) and dashed lines (100\,Myr: upper panel; 70\,Myr: lower panel) indicate two typical times when the UV ratio starts to increase and decline. }
\label{fig:uv_ratio}
\end{figure*}

To quantify the temporal evolution of the UV flux, we compute the UV flux of the UV-upturn from the Lyman break 1216\,$\rm \AA$ to 2500\,$\rm \AA$. We obtain the UV flux ratio by dividing the UV flux by the integrated flux of the entire cluster. The evolution of the UV flux ratio is presented in Figure~\ref{fig:uv_ratio}. 

For cluster ages younger than roughly 4\,Myr, the UV flux mainly originates from massive stars, and the UV flux ratio remains mostly constant in all models, with metal-rich models reaching high values. After 4\,Myr (vertical black solid line), HeMS stars start to form, and become a major contributor to the UV flux. Therefore, the UV flux continues to increase. At about 10--20\,Myr, the UV flux reaches a plateau. During the plateau period (which lasts about 70--80\,Myr, between the vertical solid and dashed lines), WDs start to form and emit in the far-UV. Although WDs continue to form and emit UV radiation, they also gradually cool down. After $\sim$70--100\,Myr, the UV flux ratio declines, and the metal-poor models surpass the metal-rich models in UV brightness, since the luminosity of metal-poor WDs is brighter \citep{althaus2015}. 

Note that we have not included helium enrichment in the stellar evolution modelling. Observational studies have found that metal-rich clusters with helium enhancement can produce very blue stars, e.g., EHB stars \citep{lee2005}. 

\section{Photometric features} \label{sec:phot}

\subsection{Color-magnitude diagrams}\label{sec:cmd}

\begin{figure*}[tb!]
\centering
\includegraphics[angle=0, width=1.\textwidth]{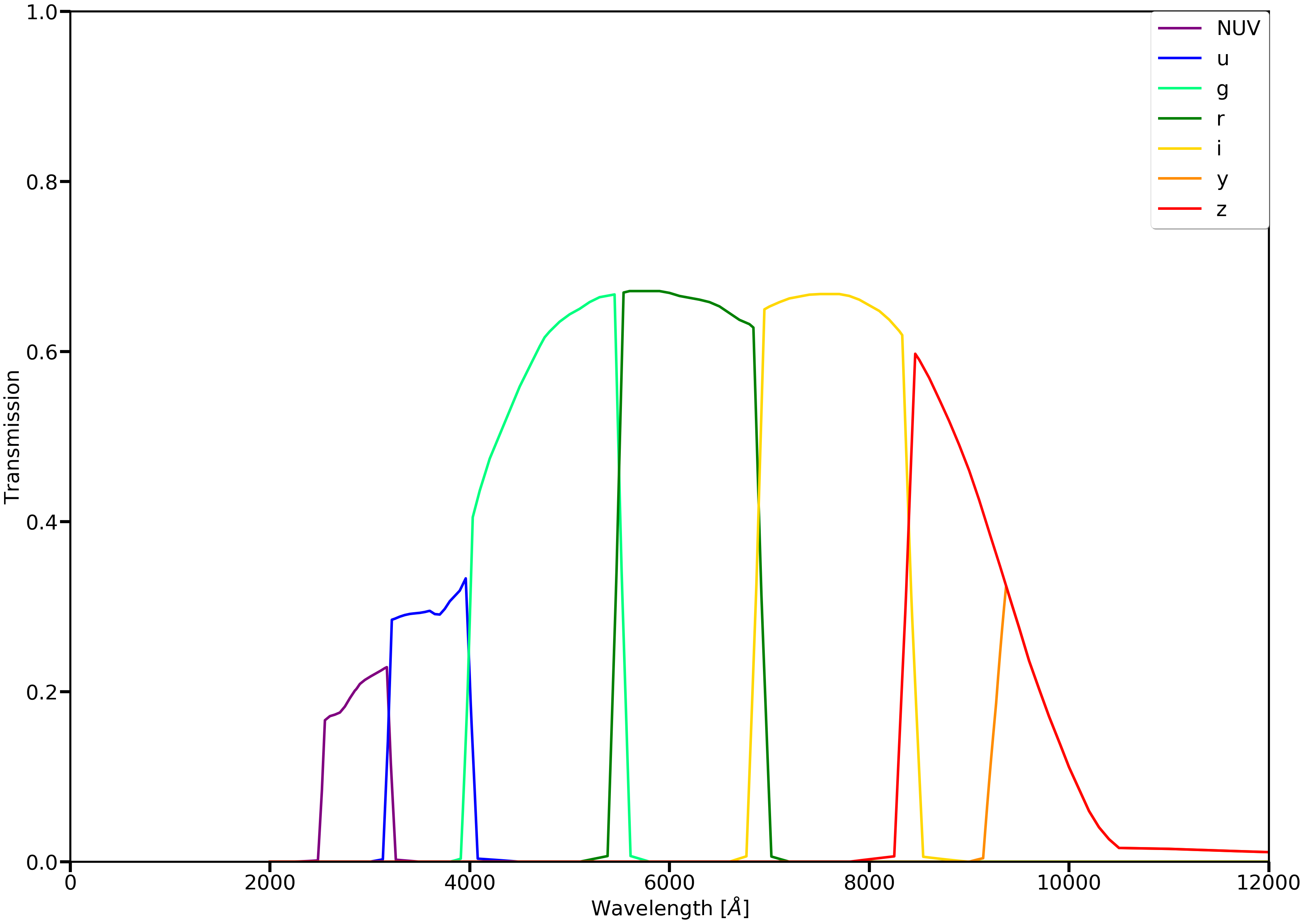}
    \caption{The transmission curves of seven filters of CSST (indicated in the top-right corner of the figure), from UV to near-infrared.}
\label{fig:CSST_filter}
\end{figure*}

\begin{figure*}[tb!]
\centering
\includegraphics[angle=0, width=1.\textwidth]{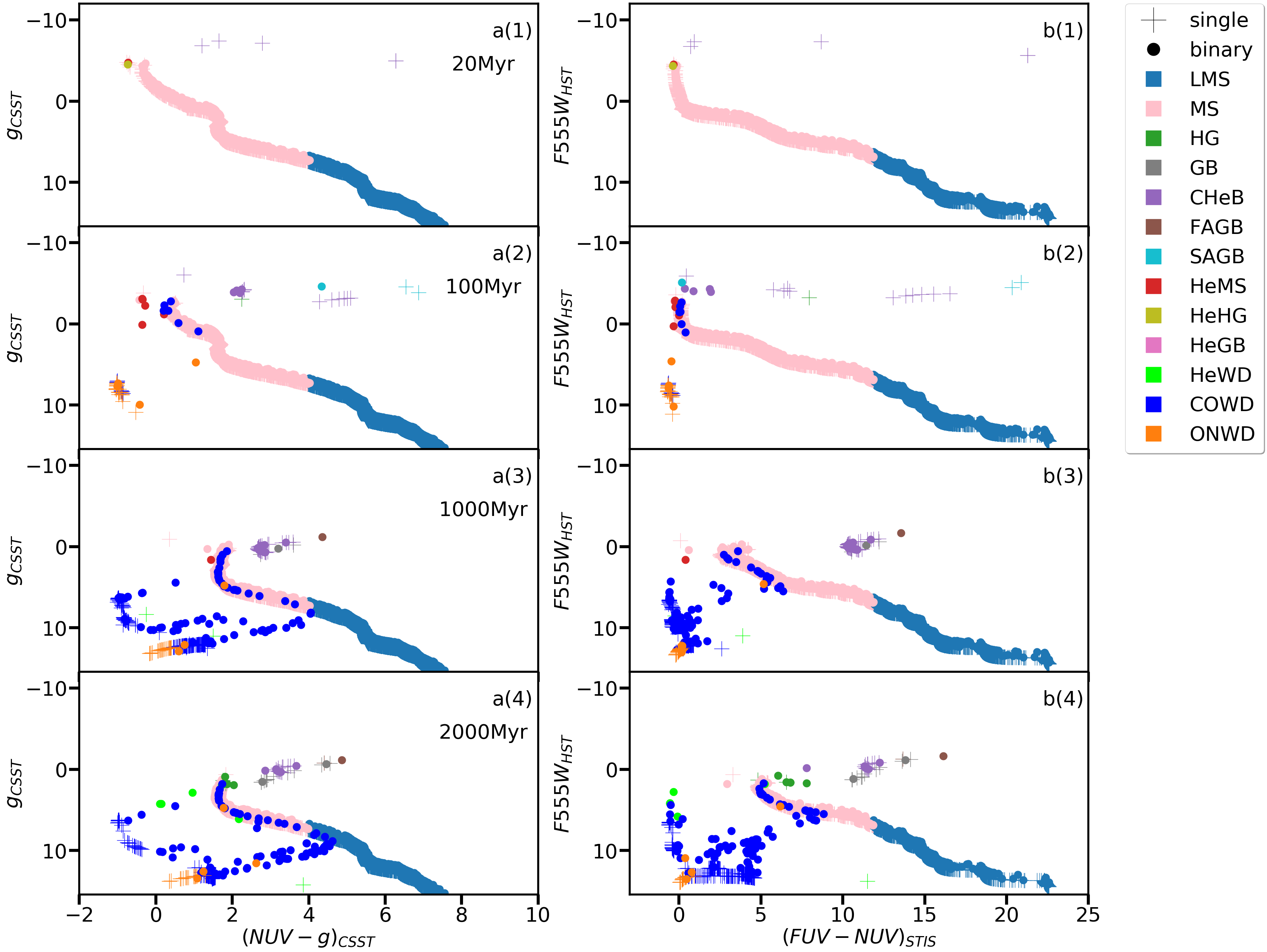}
    \caption{The color-magnitude diagram for the 10k-2Z model. The filled colored circles denote  binary systems, and the crosses denote single stars. Different symbol colors indicate different stellar types. The discontinuity of the MS at the low-mass end is due to the limited number of spectral templates in \citet{lejeune1997} for effective temperatures below 3000\,K.}
\label{fig:cmd_10k-2z}
\end{figure*}

\begin{figure*}[tb!]
\centering
\includegraphics[angle=0, width=1.\textwidth]{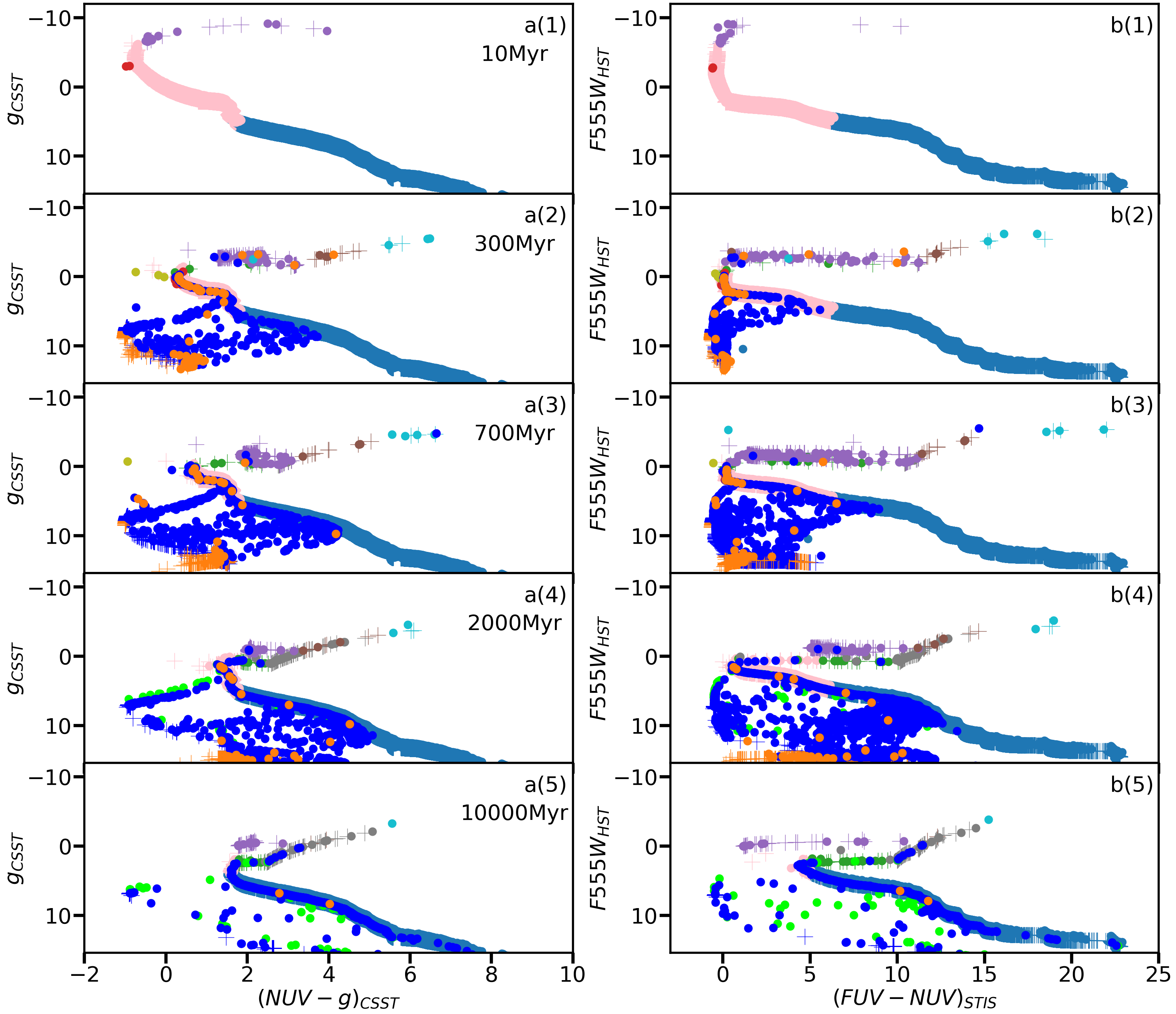}
    \caption{Color-magnitude diagrams for the 100k-4Z model. Symbols and colors are identical to those in Figure~\ref{fig:cmd_10k-2z}.}
\label{fig:cmd_100k-4z}
\end{figure*}

With the magnitudes produced by \texttt{GalevNB}, we obtain CSST and HST/STIS magnitudes for the simulated star clusters.
 The filter response curves of the CSST filters are presented in Figure~\ref{fig:CSST_filter}. 
The color-magnitude diagram (CMD) for models 10k-2Z and 100k-4Z are shown in Figures~\ref{fig:cmd_10k-2z} and~\ref{fig:cmd_100k-4z}, respectively. CMDs of other models are presented in Appendix~\ref{sec:apx_cmd}. Crosses represent single stars, while circles represent binary systems. The HeMS (red circles) are located at the left of the MS turn-off (MSTO). They occupy a region similar to the blue stragglers  or EHBs. Newly-born ONWDs are particularly hot and blue (orange circles). Their magnitude is 5\,mag fainter than the MSTO in the CSST $g$-band or HST $F555W$-band, but 3\,mag bluer in $NUV-g$ and 2--5\,mag bluer in $FUV-NUV$. 

Generally, the WD binaries are brighter than single WDs by several magnitudes. WD binaries are extremely blue at $FUV-NUV$ until the end of the simulation. Before the first 1\,Gyr, COWD+MS binaries are dominant {among WD binaries}. After 2\,Gyr, Helium WD+MS binaries start to populate, and become dominant in the population of WD binaries especially at very old ages (10\,Gyr). When a WD has a MS companion of comparable mass, the binary is located between the MS and the WD sequence. Observations found accretion and outflow in several of such these binaries, which are known as cataclysmic variable stars (CVs). The MS companion in a CV typically has a spectral type ranging between G8 and M6, or is a brown dwarf companion \citep{warner1995,knigge2011}. The MS and WD in a CV are typically separated by several solar radii. The donor star (the secondary) undergoes Roche-lobe overflow, and transfers material to the white dwarf through the inner Lagrangian point.  Note that the stellar evolution model implemented in \textsc{petar} is taken from \citet{hurley2002}; we do not model accretion disks.  


\subsection{Radial color gradients}\label{sec:color_grad}

\begin{figure*}[tb!]
\centering
\includegraphics[angle=0, width=1.\textwidth]{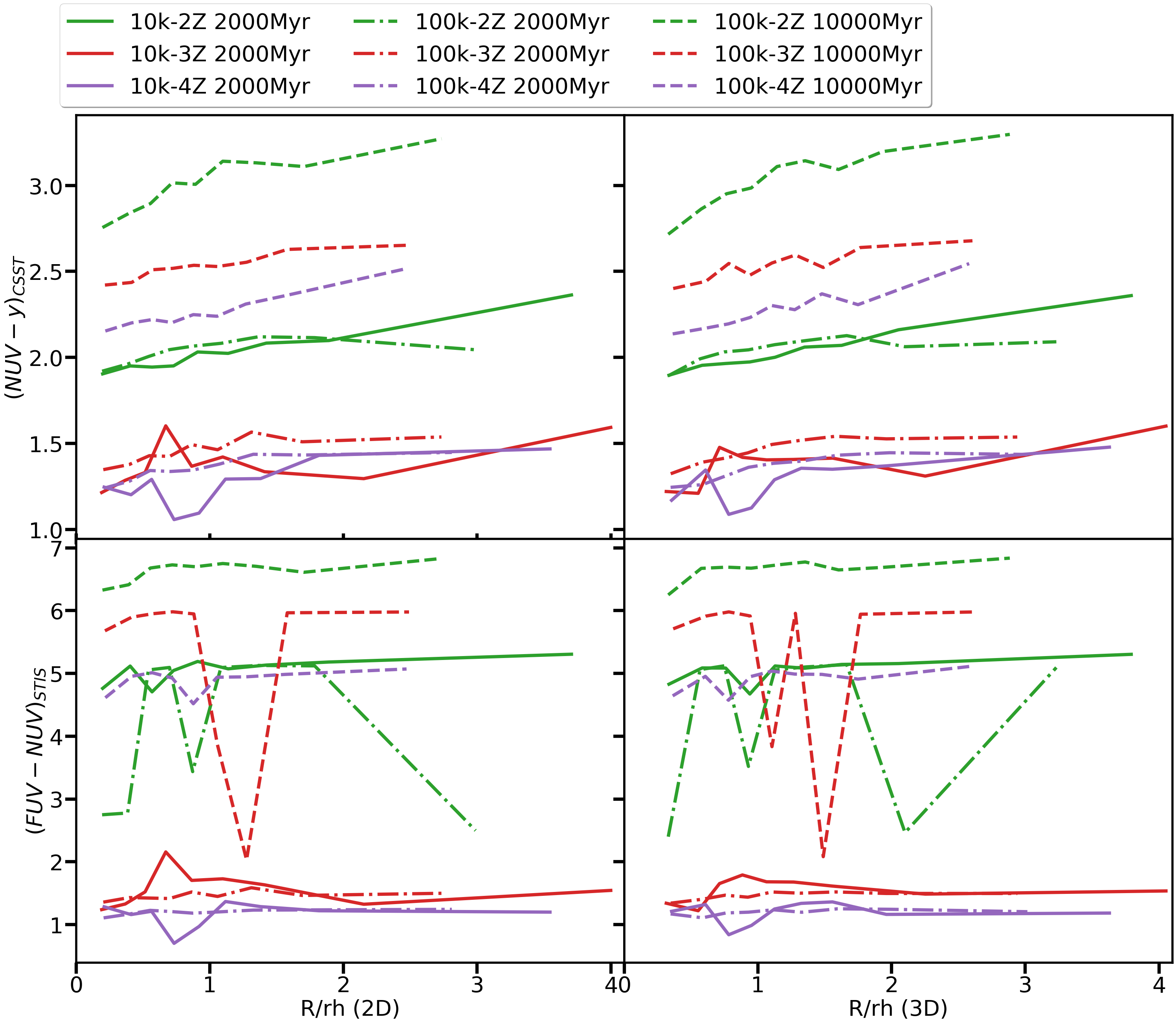}
    \caption{The dependence of the color $NUV-y$ (CSST, upper panels) and $FUV-NUV$ (HST/STIS, lower panels) on 2D and 3D cluster-centric distance (in units of the half-mass radius) for all models. All bins contain 10 percent of the total number of stars in the cluster.}
\label{fig:color_grad}
\end{figure*}

\citet{shu2021} identified a radial color gradient in the DRAGON simulations \citep{wang2016}, with bluer colors towards the cluster center and redder colors in the outskirts. This is a result of dynamical mass segregation in the star cluster. In the million-particle DRAGON simulations, a color difference of 0.2\,mag in CSST $NUV-y$ and 0.1\,mag in HST $F330W-F814W$ has been observed \citep{shu2021}. 

We obtain the integrated color of the CSST $NUV-y$ and HST/STIS $FUV-NUV$ for stars in different concentric annuli in all models  at 2\,Gyr and 10\,Gyr (i.e., at the end of the simulations). Since giant stars generate significant color fluctuations, we only include MS stars in the computation of integrated color of the clusters.  In Figure~\ref{fig:color_grad} we display the dependence of the integrated color on the cluster-centric distance. To minimize statistical fluctuations, all bins contain an equal fraction (10 percent) of the total number of stars in the cluster. Considering that individual stellar distances in remove star clusters are difficult to determine observationally, we provide both 2D and 3D cluster-centric distances.

A positive color gradient is observed in all 10k models at 2\,Gyr, and in the 100k models at both 2\,Gyr and 10\,Gyr in CSST $NUV-y$ color, from the center up to 3--4 half-mass radii ($r_h$; upper panels in Figure~\ref{fig:color_grad}). 
This positive color gradient is similar to that found in the DRAGON simulations \citep{shu2021}. At 2\,Gyr, the color gradient has a difference in $NUV-y$ of 0.2--0.3\,mag up to 3--4\,$r_{h}$ in both the 10k and the 100k models. The color gradient steepens over time, and has a color  difference in $NUV-y$ of 0.3--0.5\,mag at 10\,Gyr. The steepening of the positive color gradient in older clusters directly reflects the mass segregation. During the relaxation process, massive stars sink to the cluster center while low-mass stars migrate to the outskirts, which has been observed in Galactic star clusters  \citep[e.g.,][]{hillenbrand1998,pang2013,tang2018}. As a cluster grows older, its degree of mass segregation increases. This evolution in mass segregation has been observed in star clusters of different ages after the release of Gaia DR\,2  \citep[e.g.,][]{tang2019,pang2020,bhattacharya2021,pang2021,ye2021}. Therefore, the cluster center is bluer than its outer parts, with a higher fraction of high-mass stars. 

 However, in HST $FUV-NUV$ color (lower panels of Figure~\ref{fig:color_grad}), the fluctuations are large, especially in the 100k models. The $FUV-NUV$ color is significantly affected by the presence of blue stragglers (BSs) that are more luminous and bluer than the MSTO. 
 These are binary stars or stellar merger products. The BSS in our simulations are 2--3\,mag bluer in $FUV-NUV$ color and 3\,mag brighter in the $F555W$ band than the MSTO stars (see Figures~\ref{fig:cmd_10k-2z} and~\ref{fig:cmd_100k-4z}). They generate a minor peak in the FUV spectra (Figure~\ref{fig:sed_100k-3z}) and can be potential contributors to the UV-excess. 
There are more BSs in the 100k models since they contain more binary stars, consistent with the observations that BSs are more numerous in higher-mass and older clusters \citep{jadhav2021}. 

When we additionally remove the BSs, the color gradient in $FUV-NUV$ will be similar to that in $NUV-y$. Note that we exclude evolved stars in the integrated color derivation, which will elevate the color gradient. When all stellar populations are included, such as red giants, the color gradient suffers from large fluctuations.

\section{Dynamical evolution in the star clusters}\label{sec:dyn}

\subsection{Binary evolution}\label{sec:binary_evol}

\begin{figure*}[tb!]
\centering
\includegraphics[angle=0, width=1.\textwidth]{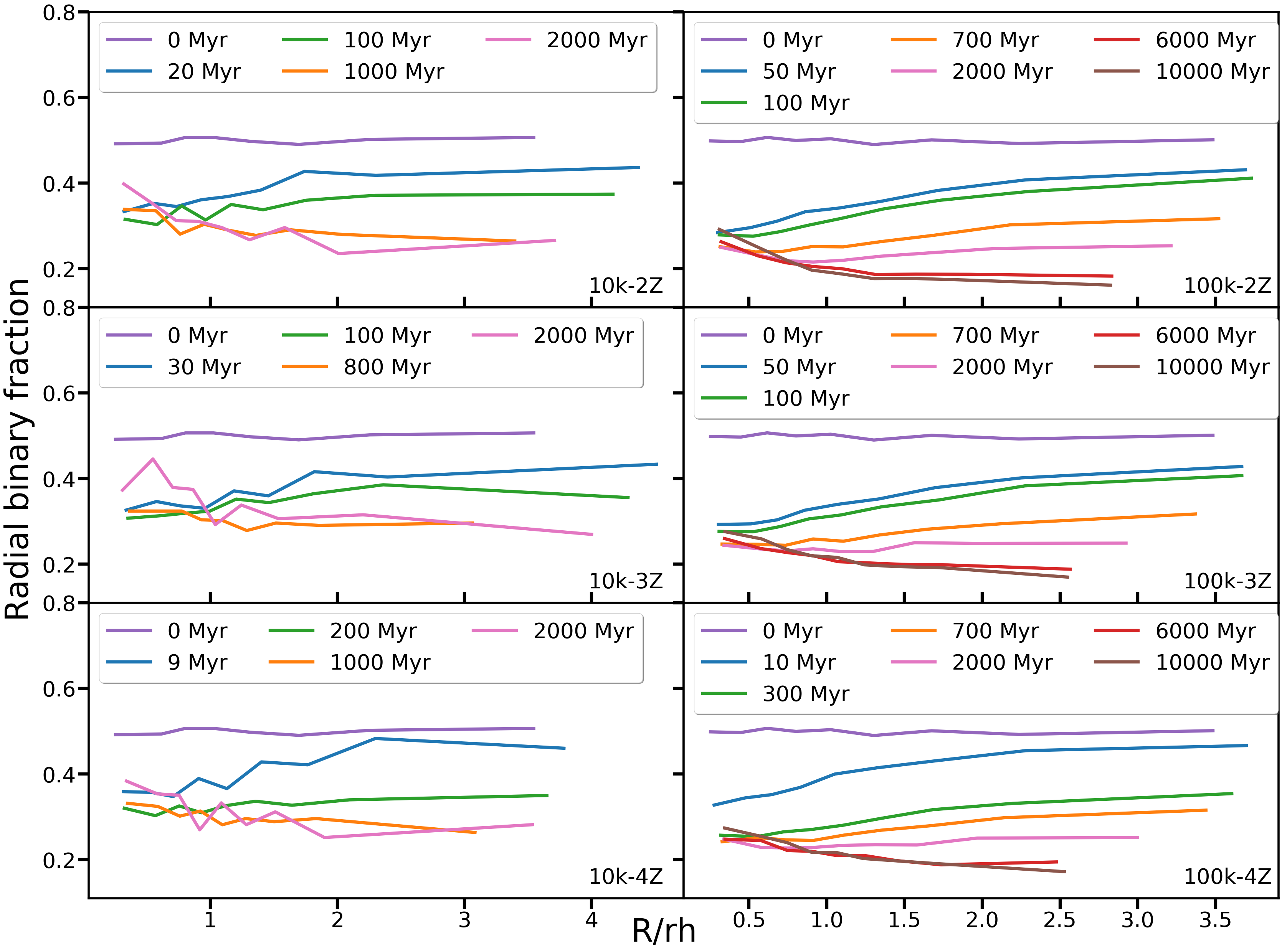}
    \caption{The evolution of the binary fraction along the radial direction for all six models. The binary fraction is computed as the fraction of binary systems among all stars (single+binary) in each annulus. Curves of different colors represent different ages, which are consistent with the SEDs (Figures~\ref{fig:cmd_10k-2z} and~\ref{fig:cmd_100k-4z}, and Figures~\ref{fig:cmd_10k-3z}, \ref{fig:cmd_10k-4z}, \ref{fig:cmd_100k-2z}, and~\ref{fig:cmd_100k-3z}) and the CMDs (Figures~\ref{fig:sed_10k-2z} and~\ref{fig:cmd_100k-4z}, and Figures~\ref{fig:cmd_10k-3z}, \ref{fig:cmd_10k-4z}, \ref{fig:cmd_100k-2z} and~\ref{fig:cmd_100k-3z}). }
\label{fig:binary}
\end{figure*}

Binary stars contribute to a large fraction of the UV flux in our simulations (see Sections~\ref{sec:SED} and~\ref{sec:cmd}). They are the major candidates for the origin of observed UV-excess in star clusters, especially WD binaries. Benefiting from direct $N$-Body simulations, we are able to follow the dynamical evolution of each of the binary candidates that contribute to the UV excess. 

Binaries are strongly affected by the long-term dynamical processes in the star cluster. This process is faster with a higher local stellar density. Therefore, the fraction of binary stars in the cluster center will differ from that in the outskirts.
We show the radial binary fraction in Figure~\ref{fig:binary}. The radial binary fraction is computed as $B/(S+B)$, where $B$ is the number of binary systems, and $S$ the number of single stars in each annulus. For all models, the binary fraction is initially uniformly assigned at all radii. During the first 100--200\,Myr, the binary fraction drops in the center and increases towards the outer parts of the cluster. 
A large number of wide binary stars are disrupted as a consequence of stellar interactions in the dense cluster center, especially during the core collapse phase. 
 The core collapse phase terminates at  
 100\,Myr for the 10k models and at 200\,Myr for the 100k models. 
 According to the Heggie-Hills law \citep{heggie1975,Hills1975}, wide binaries are easily disrupted during close encounters, while tight binaries tend to become tighter after interactions and can survive for a long periods of time.
 The boundary between these wide and tight binaries (i.e., the critical semi-major axis) is determined by the local velocity dispersion. Consequently, a higher fraction of wide binaries are disrupted in the core than in the halo of a star cluster.
 
 Subsequently, the radial binary fraction trend reverses. The binary fraction decreases significantly with increasing radius. This radially decreasing trend agrees with the results of the DRAGON simulations \citep{shu2021}. The trend grows steeper with time and becomes most profound at the end of each simulation, which is consistent with observations of GCs \citep{milone2016}.
 

\subsection{Dynamical features of WD binaries} \label{sec:CV}

\begin{figure*}[tb!]
\centering
\includegraphics[angle=0, width=1.\textwidth]{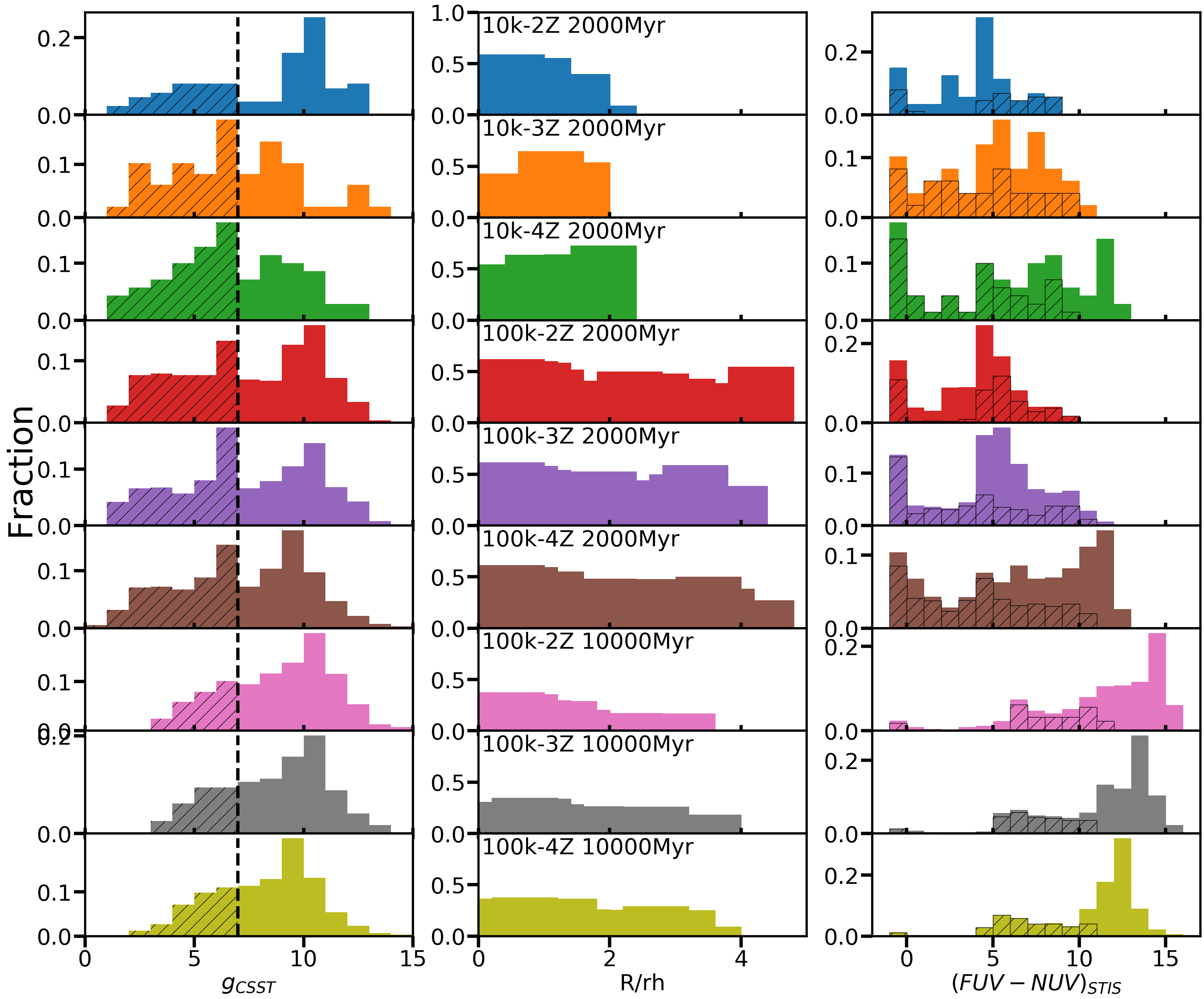}
    \caption{The properties of CV candidates at different cluster ages. The left-hand panels show the distribution of the $g_{CSST}$ magnitude of the CV candidates. The vertical dashed line highlights the brighter population of CV candidates ($g_{CSST}\lesssim7$\,mag; shaded histograms). The middle panels present the distribution of the fraction of bright CVs among all CVs in each annulus in the units of half-mass radius $\rm r_h$. The right-hand panels are the histograms of $FUV-NUV$ colors. The shaded histograms in the middle panels and right-hand panels correspond to the brighter population. }
\label{fig:CV}
\end{figure*}

In the CMDs (Figures~\ref{fig:cmd_10k-2z} and~\ref{fig:cmd_100k-4z}), WD binaries are outstandingly blue or ultraviolet, and are located between the MS and the WD sequence.  Unfortunately, the accretion algorithm for binary stars of \citet{hurley2002} may not be sufficiently accurate to model the mass-transfer processes in CV candidates. Instead, we identify CV candidates as a binary composed of a WD and a MS star. These are excellent candidate sources for the UV-excess in star cluster, consistent with the findings from our simulations in which WD binaries are a major contributor to the emitted UV radiation. We are motivated to investigate the dynamical evolution of the CV candidates.

Several studies have found that the luminosity function of CVs tends to be different for different GCs \citep{cohn2010,rivera2018,lugger2017}. 
The magnitude distributions of two GCs, NGC\,6397 and NGC\,6752, are bimodal \citep{cohn2010}. \citet{cohn2010} suggested that the optical emission of the bright group of CVs in NGC\,6397 originates from the donor stars or  accretion disks, and that of the fainter group originates from the WDs. The magnitude distribution of 47~Tuc, on the other hand, shows a unimodal distribution \citep{rivera2018}. There are fewer bright CVs in 47~Tuc per unit mass than in the other clusters. The cumulative radial distributions of bright CVs in both NGC\,6752 and NGC\,6397 show a strong concentration towards the cluster centre, while the fainter CVs tend to be located at larger radii \citep{rivera2018,lugger2017}. 

In order to be able to compare our findings with observations, we analyze the $g_{CSST}$ magnitude and cluster-centric distance for the CV candidates in our models (see Figure~\ref{fig:CV}). We find that only within 2\,Gyr, CV candidates have a bimodial distribution in $g_{CSST}$ for both the 10k models and the 100k models. There is a peak around $g_{CSST}\sim10$\,mag and another peak around $g_{CSST}\sim7$\,mag. Note that the magnitudes computed from the $N$-body models are absolute magnitudes. As the cluster grows older at 10\,Gyr (100k simulations), only the $g_{CSST}\sim10$\,mag are visible, with a plateau distribution at the bright region. We select bright CV candidates, i.e., those  with magnitudes brighter than $g_{CSST}=7$\,mag (bright population, shaded histogram in the left-hand panels of Figure~\ref{fig:CV}), and highlight the fraction of bright CVs among all CV candidates in the histogram of cluster-centric distance (middle panels of Figure~\ref{fig:CV}). 

During the first 2\,Gyr the fraction of bright CVs remains almost constant from the cluster center to the outskirts in all models. We do not observe a preference for the bright CVs to be located in the cluster centre at this stage. When the clusters grow as old as 10\,Gyr, the fraction of bright CVs in the center is marginally higher than in the outskirts. This trend is most pronounced in the 100k-2Z model. Although bright CV candidates do not show a spatial preference, they are indeed much brighter in $FUV-NUV$ (right-hand panels of Figure~\ref{fig:CV}), and mainly occupy the peak at $FUV-NUV=0$\,mag. At 10\,Gyr, the bright CV candidates are solely brighter in  $FUV-NUV$, compared to the faint population.


These bright CV candidates are COWD and HeWD binaries (see Figures~\ref{fig:cmd_10k-2z} and~\ref{fig:cmd_100k-4z}). This may be an evolutionary effect. The progenitors of COWD and HeWD are intermediate-mass and low-mass MS stars. They require a longer time than higher-mass stars to experience mass segregation, since the mass segregation timescale is inversely proportional to stellar mass \citep[e.g.,][]{pang2013}. At 2\,Gyr, these MS stars are segregated into the cluster center and finally become WDs. As the cluster grows older, the brightness of COWD or HeWD declines and dynamically diffuse to larger radii as a consequence of interactions with neighbouring single stars or binaries. Hence, the old CV candidates show no peak at the bright magnitude or in cluster-centric distance. Finally, the bright CV candidates are major source of $FUV$ radiation in the cluster. 

Therefore, the location of bright CV candidates closer to the cluster center may be due to internal dynamical evolution of the cluster. However, we cannot exclude that their brightness may originate from mass transfer between CV components, as suggested in earlier observational studies \citep{rivera2018,lugger2017}. Limited by the binary evolution model in \textsc{petar}, detailed simulation of CV candidates with realistic analytical models in star cluster environment is necessary in the future.

\section{Discussion}
\label{sec:dis}

\subsection{Comparison to observations}\label{sec:comp}

\begin{figure*}[tb!]
\centering
\includegraphics[angle=0, width=1.\textwidth]{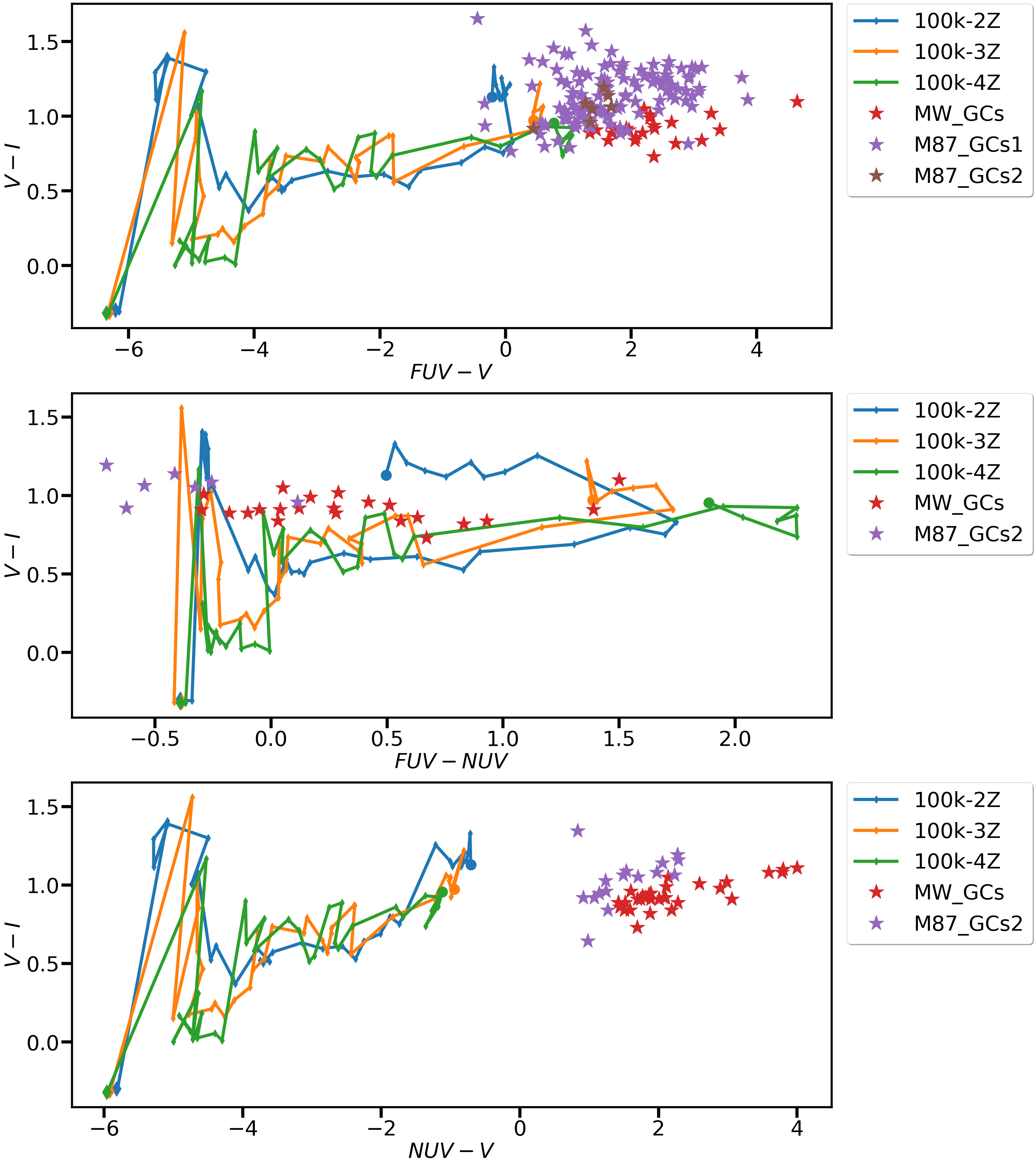}
    \caption{Color-color diagrams in HST/STIS color: {\em upper panel}: $FUV-V$ vs. $V-I$; {\em middle panel}: $FUV-NUV$ vs. $V-I$; {\em lower panel}: $NUV-V$ vs. $V-I$. Only the 100k models are presented. The simulation starts at the diamond (0\,Myr) and ends at the filled circles (10\,Gyr). Star-symbols are observed GCs in the Milky Way (red) and in M\,87 (purple and brown). The purple stars are GCs with both $FUV$ and $NUV$ observations, while the brown stars only have $FUV$ observations. }
\label{fig:obs_comp1}
\end{figure*}

To verify the consistency of our models with observations, we compare the UV photometric features of our 100k models to those of GCs in the M\,87 and in the Milky Way (MW). Note that the 100k models only reach the lower-mass limit of GCs. \citet{sohn2006} observed UV-bright stars in old, metal-rich GCs in the giant elliptical galaxy M\,87 with HST/STIS $FUV$ and $NUV$ photometry. \citet{sohn2006} recomputed the UV photometry of the MW GC samples from \citet{dorman1995} with the metallicity [Fe/H] and reddening $E(B-V)$ from \citet{harris1996}. We adopt the photometry in  tables~3 and~4 from \citet{sohn2006} for the M\,87 and that of table~6 from \citet{sohn2006} for the MW GCs. 

In Figure~\ref{fig:obs_comp1} we displace the color-color evolution of the 100k models, starting from a diamond symbol. At the end the simulations (filled circles), the $FUV-V$ color of our models only matches the bluest GCs in the M\,87. However, our models are bluer than all MW GC samples. The $V-I$ of the models agrees with the observed GCs in both galaxies. The 100k-2Z and 100K-3Z model overall is consistent with the $FUV-NUV$ color of MW GCs. The $FUV-NUV$ color of GCs in M\,87 is about 1--2\,mag bluer than the model predicted at 12\,Gyr. All three models predict a bluer $NUV-V$ color (1\,mag brighter), than the observed GCs. GCs in M\,87 is 0.5--1\,mag brighter in $FUV$ and $NUV$ photometry than those in the MW. Since the metallicity in M\,87 is super-solar, helium enhancement is suggested to be a promising mechanism to explain the UV excess in their GCs \citep{sohn2006}. However, in our simulation, we cannot modify the helium abundance in the stellar evolution models. Thus may induce the observed discrepancy in Figure~\ref{fig:obs_comp1}. 

\begin{figure*}[tb!]
\centering
\includegraphics[angle=0, width=1.\textwidth]{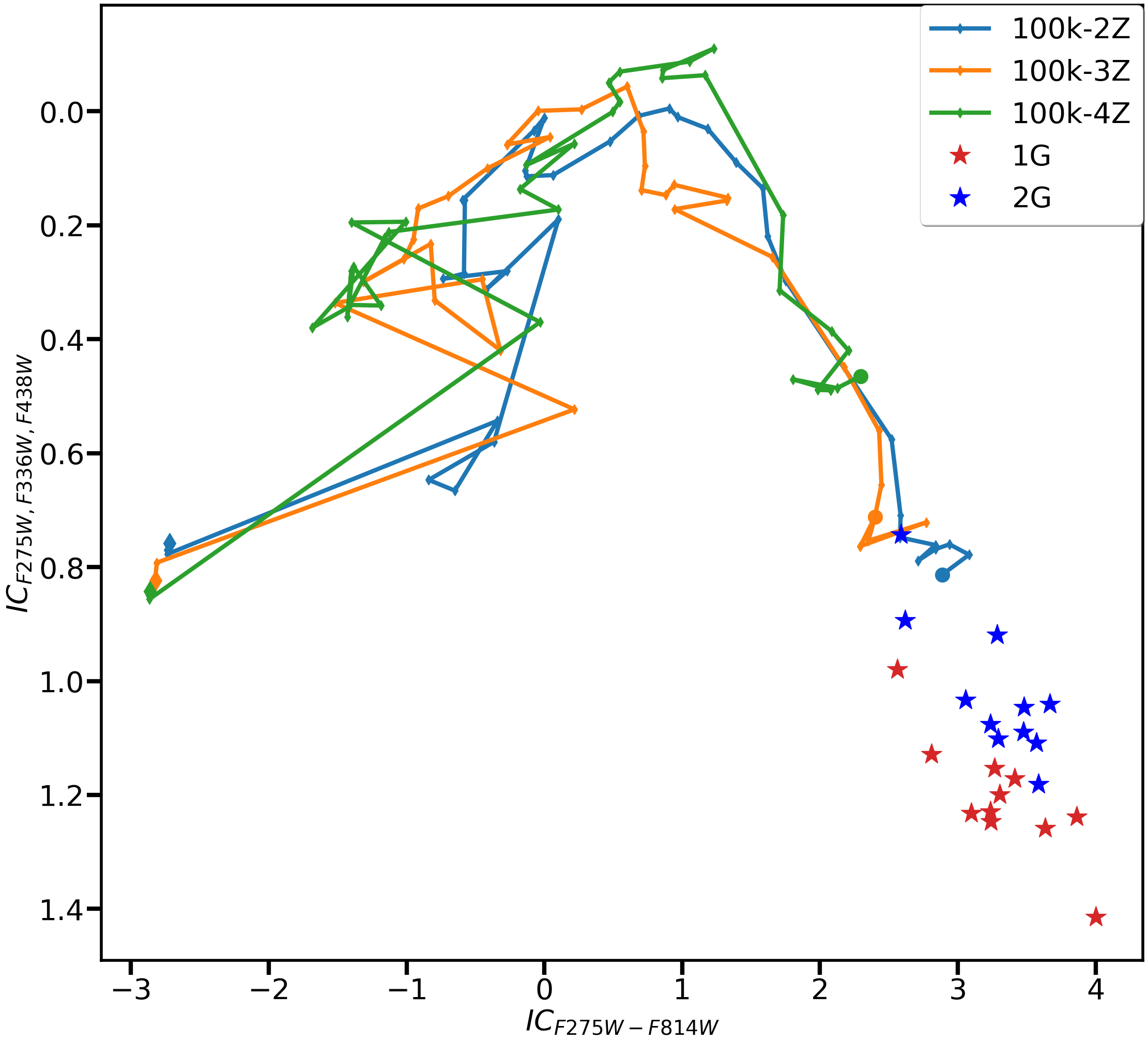}
    \caption{IC$_{F275W,F336W,F438W}$ vs. IC$_{F275W-F814W}$ integrated two-color diagram for 100k models. Data for the eleven GCs (star symbols) are obtained from \citet{jang_integrated_2021}. The red stars represent the population of first generation stars in the GCs, and the blue stars represent the second generation.}
\label{fig:obs_comp3}
\end{figure*}

\citet{jang_integrated_2021} separated the first and second generation of stars in eleven GCs using the  (m$_{F336W}$-m$_{F438W}$ vs. m$_{F275W}$-m$_{F336W}$) color-color diagram. Their results show that the integrated color of the second generation stars with helium enrichment is bluer in the color IC$_{F275W,F336W,F814W}$ \citep{milone2015} than the first generation stars. We display the evolutionary tracks of the three 100k models in Figure~\ref{fig:obs_comp3}. At an age of 12\,Gyr, the simulated clusters 
match the integrated color IC$_{F275W,F336W,F814W}$ and IC$_{F275W,F814W}$ at the bluest part of the second generation stars (blue stars in Figure~\ref{fig:obs_comp3}), but are about 0.2--0.4\,mag bluer than the first generation stars (red stars in Figure~\ref{fig:obs_comp3}) in IC$_{F275W,F336W,F814W}$. Note that all our models have a lower metallicity than the eleven observed GCs analysed in \citet{jang_integrated_2021}. Therefore, we predict a bluer color, even for first generation stars.

\subsection{Predictions for future CSST observations}\label{sec:predict}

\begin{figure*}[tb!]
\centering
\includegraphics[angle=0, width=1.\textwidth]{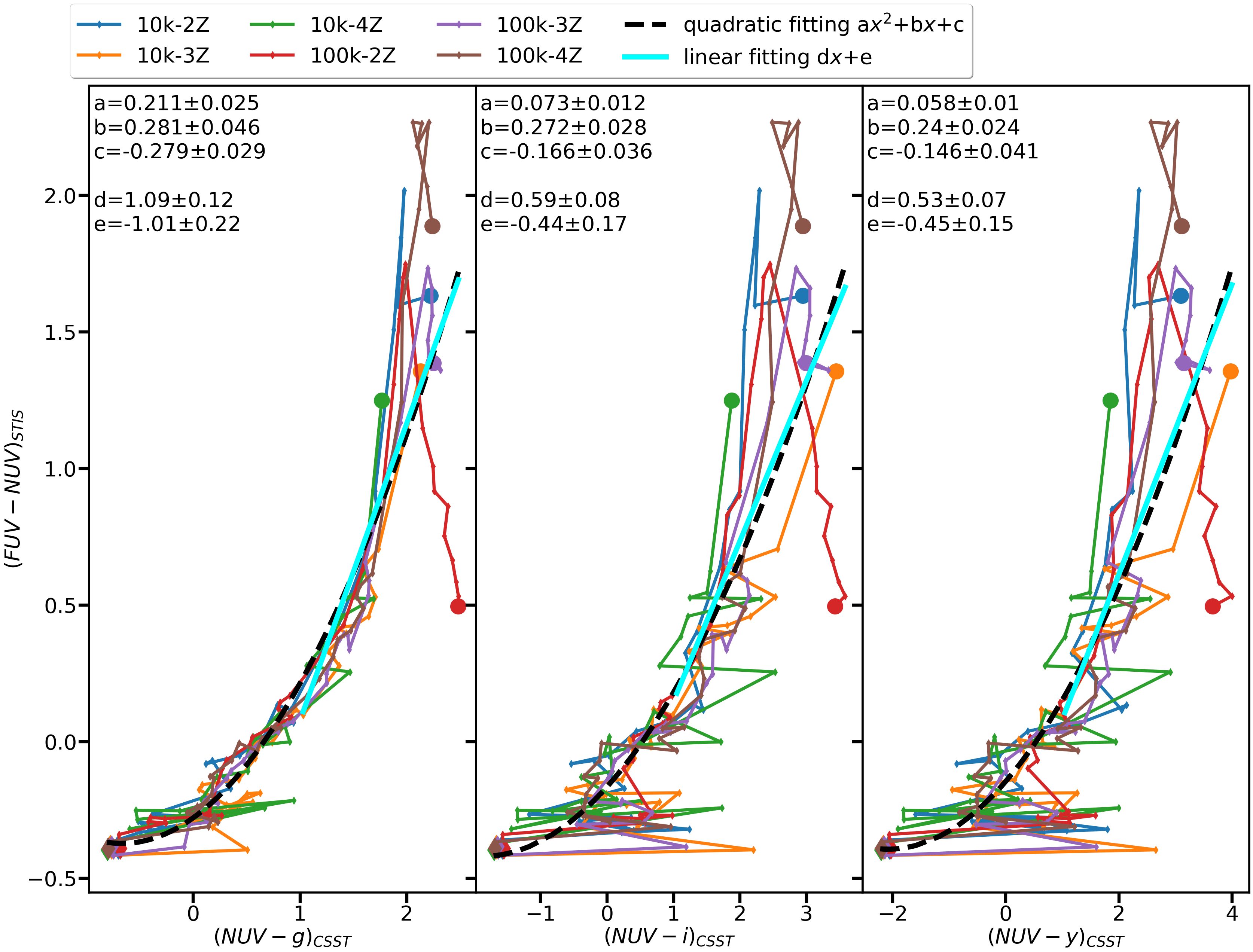}
    \caption{Correlation between $FUV-NUV$ (HST/STIS) and $NUV-g$ (CSST) for all models. The dashed black curve is a least-square fit to all simulated data points. Each point represents one snapshot of the simulation. The simulation starts at the diamond (0\,Myr) and ends at the filled circles (2\,Gyr for the 10k models and 10\,Gyr for the 100k models). }
\label{fig:obs_comp2}
\end{figure*}

To observe the UV-excess in star cluster, $FUV$ filter is the best wavelength to cover the UV-upturn region. The $NUV$ band in CSST goes down to 2500\,\AA, where the UV-upturn SED just starts to rise toward shorter wavelength. Therefore, the ability and sensitivity of CSST photometry in detecting UV-excess in star cluster needs to be tested.  We present the correlation between the $FUV-NUV$ color of HST/STIS and $NUV-g$ of CSST in Figure~\ref{fig:obs_comp2}. Different colored curves represent different models. The simulated clusters begin at the lower left of the figure (diamond symbols) and evolve to the upper right corner (filled circles, end of the simulation). We fit the correlation between HST and CSST colors with a parabola, coefficients of which are shown in the upper left corner in each panel. The correlation has a larger scatter in $NUV-i$ and $NUV-y$ than in $NUV-g$. Therefore, we suggest that $NUV-g$ is the best color search for a correlation with HST/STIS $FUV-NUV$. 

 At colors bluer than $NUV-g=1$\,mag, the CSST $NUV-g$ is not sensitive to HST $FUV-NUV$. The slope of the parabola in this range is very shallow. When the color is redder than $NUV-g=1$\,mag, the correlation between $FUV-NUV$ and $NUV-g$ becomes steeper and close to linear relation, with a slope of $1.09\pm0.12$. We can use the correlation to convert future observed CSST $NUV-g$ into HST/STIS $FUV-NUV$. At $NUV-g=1$\,mag, it corresponds to 200\,Myr in simulations. 
 
Therefore, the best observational color to detect UV-excess in star clusters for the future CSST observation lies in $NUV-g>1$\,mag. In this color range, we can convert the observed integrated cluster color $NUV-g$ into HST/STIS color $FUV-NUV$ via the linear correlation:
\begin{equation}
    FUV-NUV=(1.09\pm0.12)\times(NUV-g)+(-1.01\pm0.22)
\end{equation}
We suggest that the best target clusters to look for the signature of UV-excess or UV-upturn via CSST are clusters older than 200\,Myr.

\section{Summary}\label{sec:summ}

We evolve a set of six $N$-body models of star clusters using \textsc{petar} code, with different particle numbers ($N=10$k and $N=100$k) and different metallicities ($Z=0.01$, $0.001$, and $0.0001$). We adopt simple stellar populations; we do not include multiple stellar population, and we do not include helium variations in the $N$-body models. Using the \texttt{GalevNB} package, we convert the physical stellar properties generated by \textsc{petar} into SEDs, spanning from the far UV to the near-IR, and into photometric CSST and HST magnitudes. By observing the long-term evolution of the photometric and spectral features of the simulated clusters, we identify the stars that produce the UV-excess/UV-upturn feature in the star cluster's SEDs at wavelengths between 1216\,\AA  and 2500\,\AA. We also analyze the dynamical evolution of UV-excess candidate stars: WD+MS binaries. The main objective of our study is to predict the observational features of star clusters for future CSST observations and identify the sensitivity and ability of the CSST in detecting UV-excess in star clusters. Our main results can be summarized as follows.

(1) Three types of stars are identified as candidates for UV-excess candidates in star clusters: SAGBs, naked helium stars, and WDs. Among these, SAGBs are present in the star cluster at young ages. Due to the short life time of SAGBs, their contribution to the UV-excess is not significant. On the other hand, naked helium stars continue to radiate in the FUV until $t\approx 2$\,Gyr in the massive cluster models (100k). In the long term, the greatest contributors to the UV-excess are the three types of WDs: ONWDs, COWDs, and HeWDs.

(2) The UV flux ratio (i.e., the ratio between the UV flux in the wavelength range 1216--2500\,$\rm \AA$ divided by the total flux of the cluster) in the SEDs of the six models reaches a plateau period from 10--80\,Myr. During the plateau phase, the UV flux is mainly produced by the hot and blue naked helium stars and by young WDs. 

(3) A color gradient with a CSST $NUV-y$ color difference of 0.2--0.5\,mag within 3--4\,$r_{h}$ is observed in all models. The color gradient 
becomes more pronounced in the 100k models as the clusters age from 2\,Gyr to 10\,Gyr. This is expected from mass segregation due to dynamical relaxation. On the other hand, no color gradient is found in the HST/STIS $FUV-NUV$ color. Large fluctuations in $FUV-NUV$ are induced by blue stragglers. 

(4) Soft binary systems are rapidly disrupted in the dense core region of a cluster, especially during the core collapse phase. Therefore, the radial binary fraction declines toward the cluster center within the first 100\,Myr and 200\,Myr for the 10k and 100k models, respectively. After the termination of the core collapse phase, two-body relaxation dominates the internal dynamical processes. The binary fraction increases toward the cluster center, as the process of mass segregation continues. 

(5) We select the WD+MS binaries as the CV candidates. At an age of 2\,Gyr, there is a clear evidence of a bi-modal distribution in the $g_{CSST}$ magnitude distribution of CV candidates. We do not observe a significant spatial preference for the bright population of the CV candidates during the first 2\,Gyr. They become marginally centrally concentrated at 10\,Gyr, especially for 100k-2Z model. The bright CV candidates are also especially bright in $FUV-NUV$ and are major contributors to the UV-excess features in the cluster. At the end of the simulation of the 100k models at $t=10$\,Gyr, the bright CV population faints out and their magnitude distribution becomes uni-modal. 

(6) By comparing the HST/STIS color of UV-excess GCs in M\,87 and in the MW, we find that our models agree with observations in $V-I$. The $FUV-V$ color matches the M\,87 GCs, but bluer by 0.5\,mag than the MW GCs. The $NUV-V$ is also bluer than all GC samlpes by $\sim1$\,mag.  On the other hand, the $FUV-NUV$ color is redder than that of observed GCs by up to 1--2\,mag. This discrepancy may originate from a different helium abundance in these GC populations. In order to tackle this problem, additional numerical simulations with varying helium abundances in the stellar evolution are required. 

(7) The CSST $NUV-g$ color is most sensitive to HST/STIS $FUV-NUV$ in the color range $NUV-g>1$\,mag. We find a strong correlation between the CSST $NUV-g$ color and HST/STIS $FUV-NUV$ with our models. Especially after the cluster grows older than 200\,Myr, this correlation approaches a linear relation:  $FUV-NUV=(1.09\pm0.12)\cdot(NUV-g)+(-1.01\pm0.22)$. For future CSST observations of extra-galactic star clusters, this linear correlation may provide an essential relation that allows conversion of $NUV-g$ colors into $FUV-NUV$ colors, and can help to further unravel the origins of the observed UV excess in star clusters.

\normalem
\begin{acknowledgements}
We thank the anonymous referee for advice to improve the paper. We give thanks to Prof. Chengyuan Li from Sun Yat-sen University for providing helpful comments on multiple populations. 
We are grateful to Ms. Han Qu for computing the zero points for all filters of CSST. 
We acknowledge the science research grants from the China Manned Space Project with NO. CMS-CSST-2021-A08. X.Y.P. is grateful to the financial support of National Natural Science Foundation of China, No. 12173029, and the Research Development Fund of Xi’an Jiaotong-Liverpool University (RDF-18--02--32). L.W. thanks the support from the one-hundred-talent project of Sun Yat-sen University and the National Natural Science Foundation of China through grant 12073090 and financial support from JSPS International Research Fellow (Graduate School of Science, The University of Tokyo). M.B.N.K acknowledges support from Research Development Fund project RDF-SP-93 of Xi’an Jiaotong-Liverpool University. 

\end{acknowledgements}
  
\appendix
\label{sec:apx}

\section{Spectral energy distribution}\label{sec:apx_sed}

\begin{figure*}[tb!]
\centering
\includegraphics[angle=0, width=1.\textwidth]{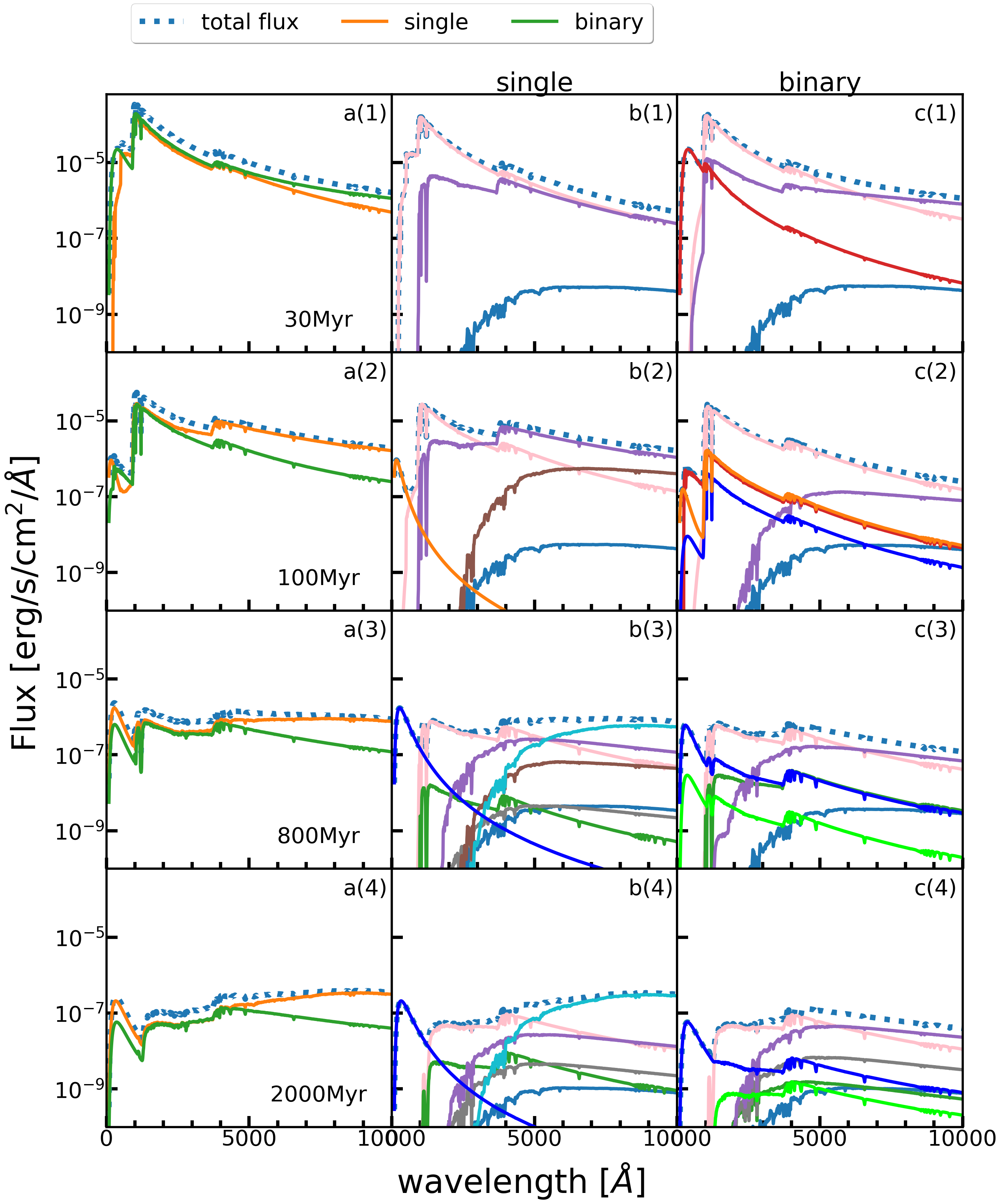}
    \caption{The SEDs of model 10k-3Z. Panels~(a) show the SEDs for the entire star cluster (dotted blue curve), the single stars (solid orange curve), and the binary systems (solid green curve). Panels~(b) and~(c) show the SEDs for individual populations of single stars and binary systems, respectively, in which we highlight the SEDs for different SPs with different color curve. The list of abbreviations for the SPs is provided in Table~\ref{table:SP}. Symbols and colors are identical to those in Figure~\ref{fig:sed_10k-2z}. }
\label{fig:sed_10k-3z}
\end{figure*}

\begin{figure*}[tb!]
\centering
\includegraphics[angle=0, width=1.\textwidth]{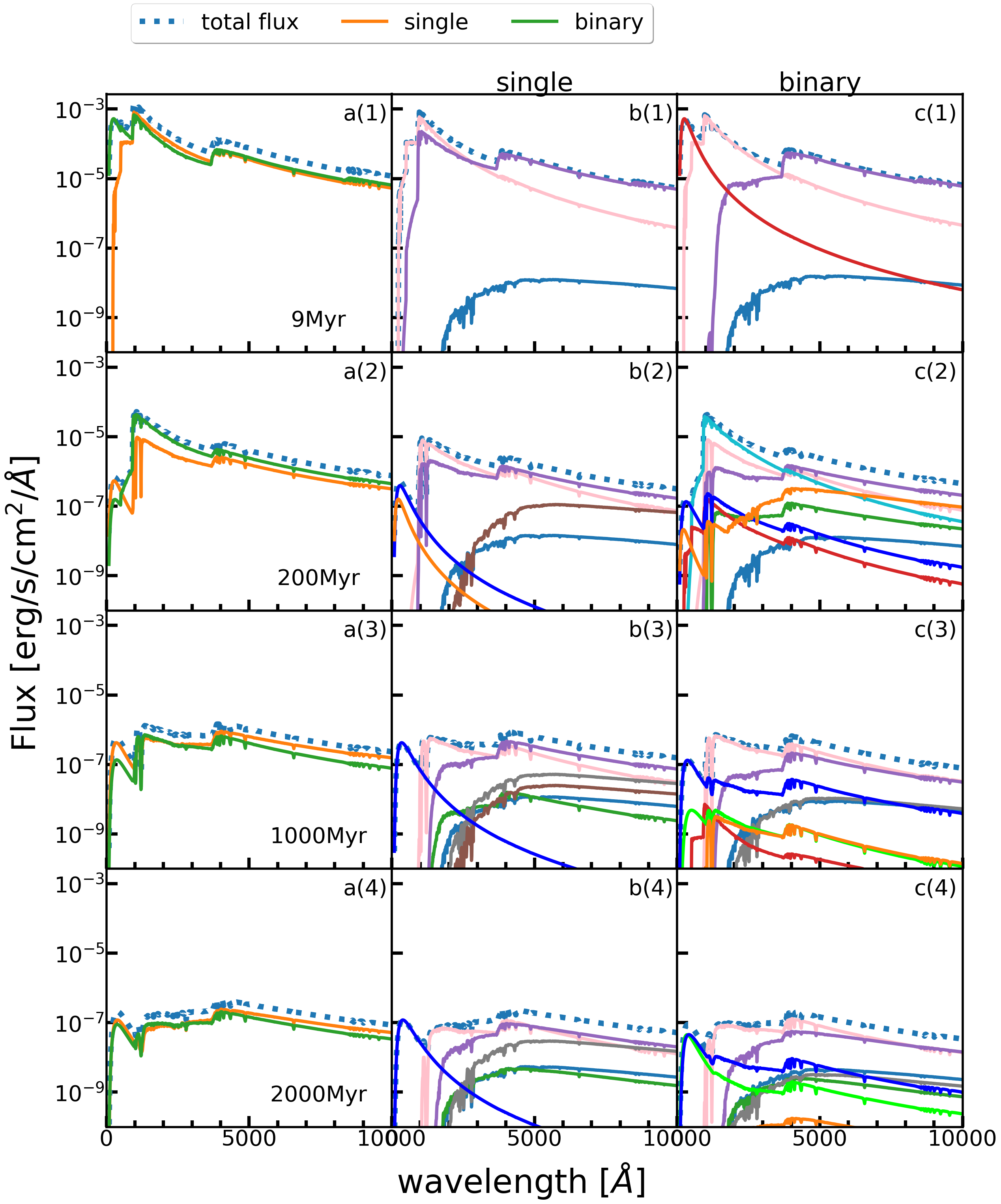}
    \caption{The SEDs for model 10k-4Z. Symbols and colors are identical to those in Figure~\ref{fig:sed_10k-2z}. }
\label{fig:sed_10k-4z}
\end{figure*}

\begin{figure*}[tb!]
\centering
\includegraphics[angle=0, width=1.\textwidth]{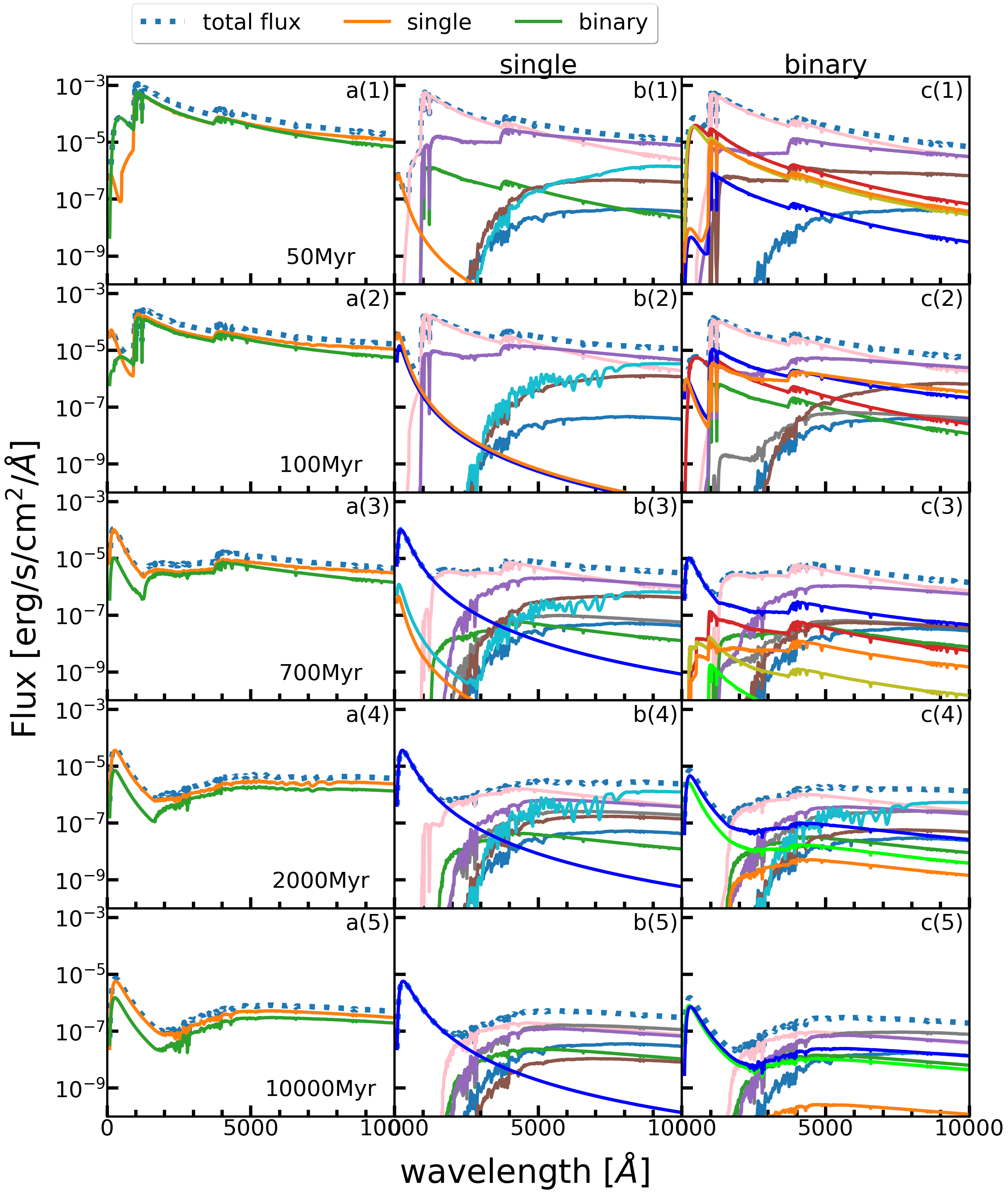}
    \caption{The SEDs for model 100k-2Z. Symbols and colors are identical to those in Figure~\ref{fig:sed_10k-2z}. }
\label{fig:sed_100k-2z}
\end{figure*}

\begin{figure*}[tb!]
\centering
\includegraphics[angle=0, width=1.\textwidth]{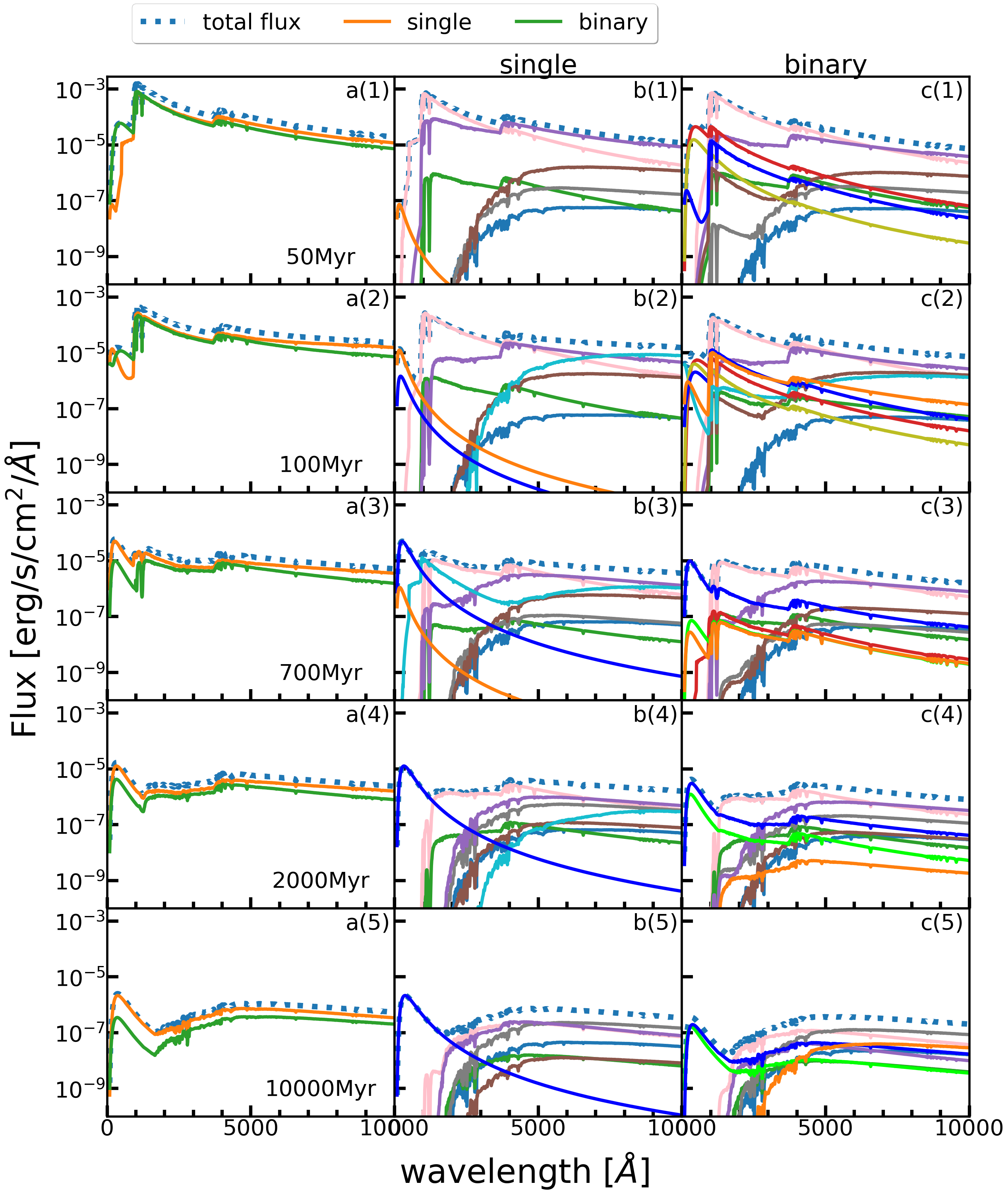}
    \caption{The SEDs for model 100k-3Z. Symbols and colors are identical to those in Figure~\ref{fig:sed_10k-2z}. }
\label{fig:sed_100k-3z}
\end{figure*}

\section{Color magnitude diagrams}\label{sec:apx_cmd}

\begin{figure*}[tb!]
\centering
\includegraphics[angle=0, width=1.\textwidth]{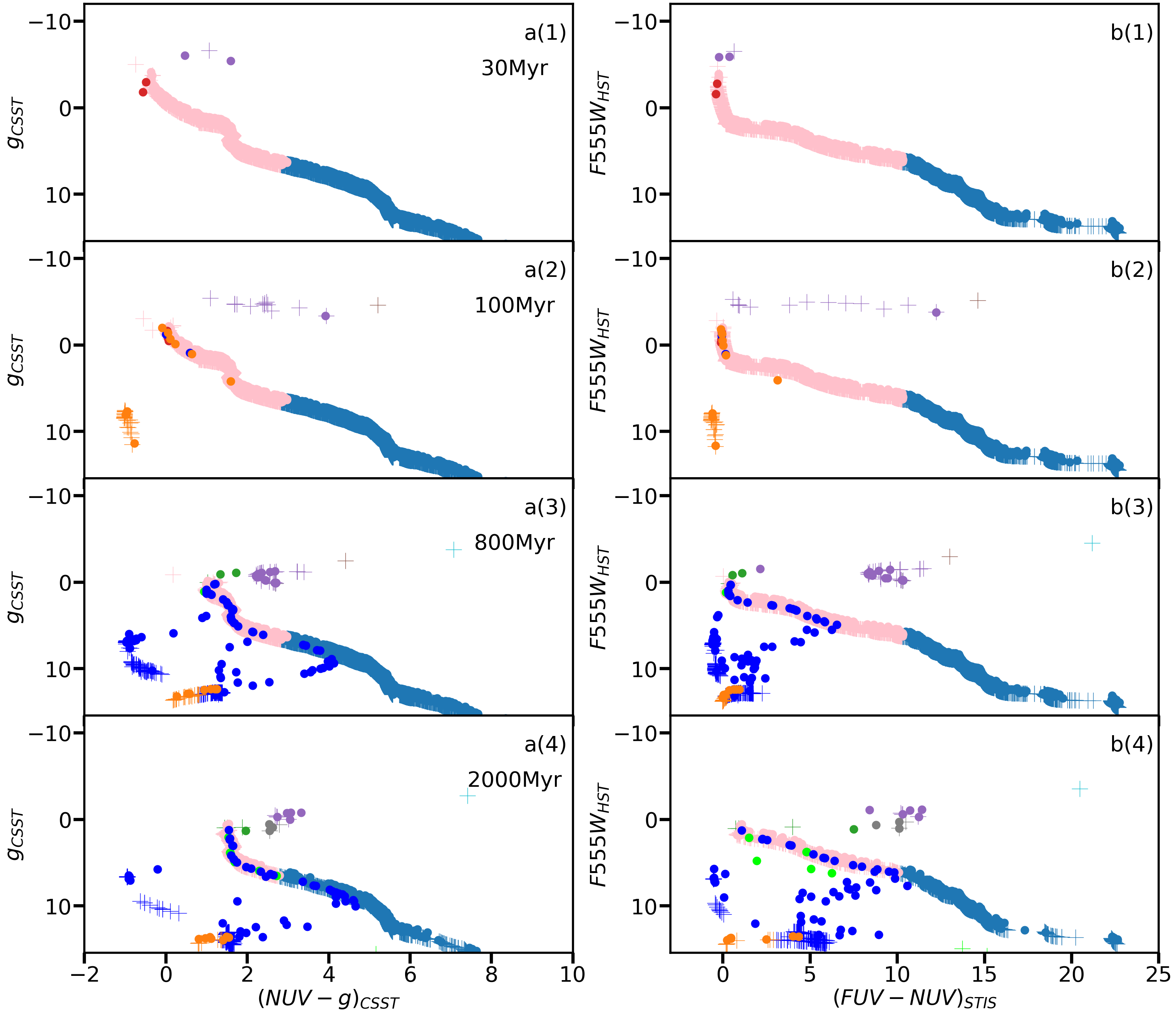}
    \caption{The CMDs for model 10k-3Z. The filled color circles denote  binary stars and the crosses denote the single stars. Different symbol colors denote different stellar types. The discontinuity of the MS at the low-mass end is due to a smaller number of spectral template in \citet{lejeune1997} for effective temperature below 3000\,K.}
\label{fig:cmd_10k-3z}
\end{figure*}

\begin{figure*}[tb!]
\centering
\includegraphics[angle=0, width=1.\textwidth]{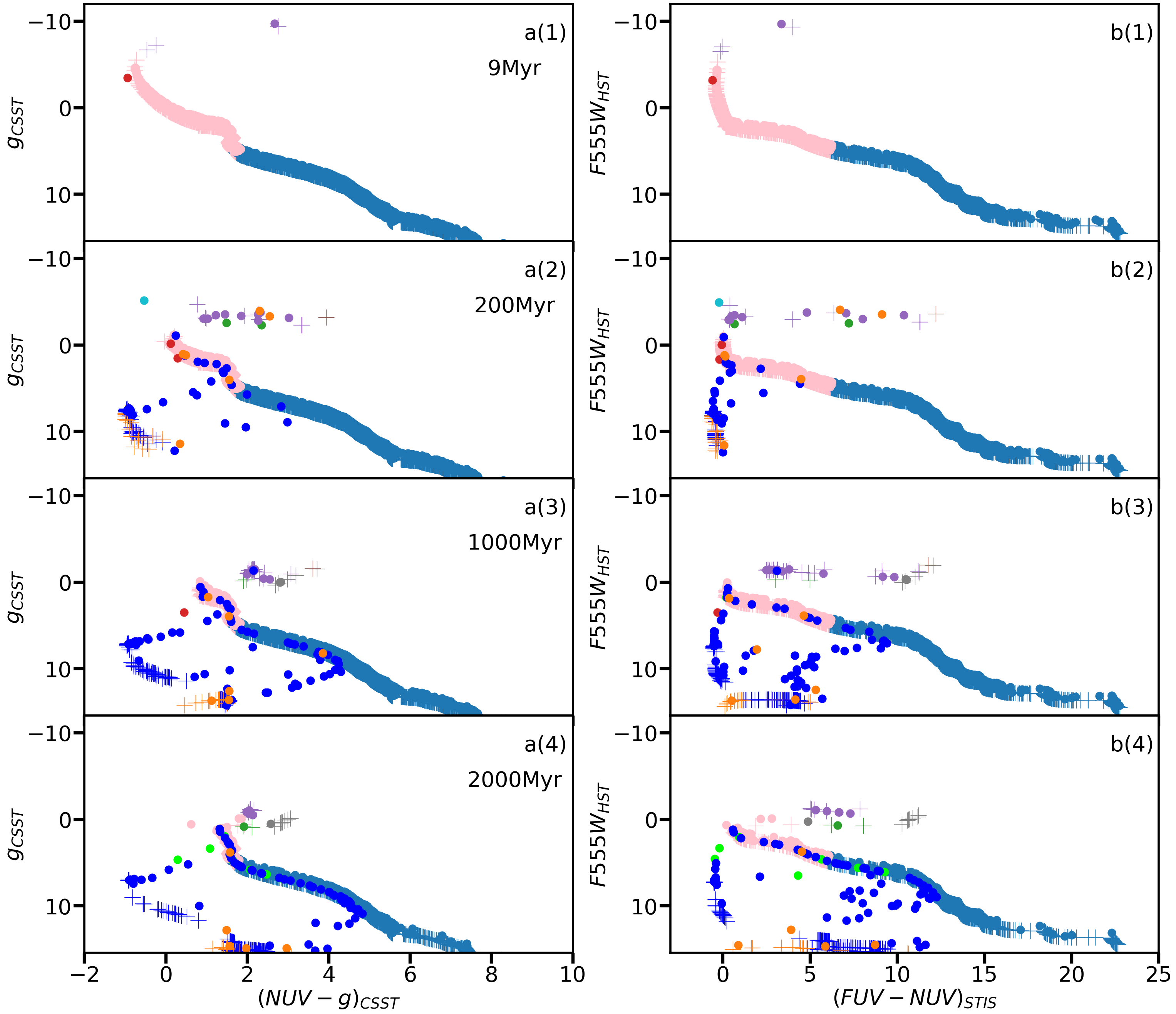}
    \caption{The CMDs for model 10k-4Z. Symbols and colors are identical to those in  Figure~\ref{fig:cmd_10k-2z}.}
\label{fig:cmd_10k-4z}
\end{figure*}

\begin{figure*}[tb!]
\centering
\includegraphics[angle=0, width=1.\textwidth]{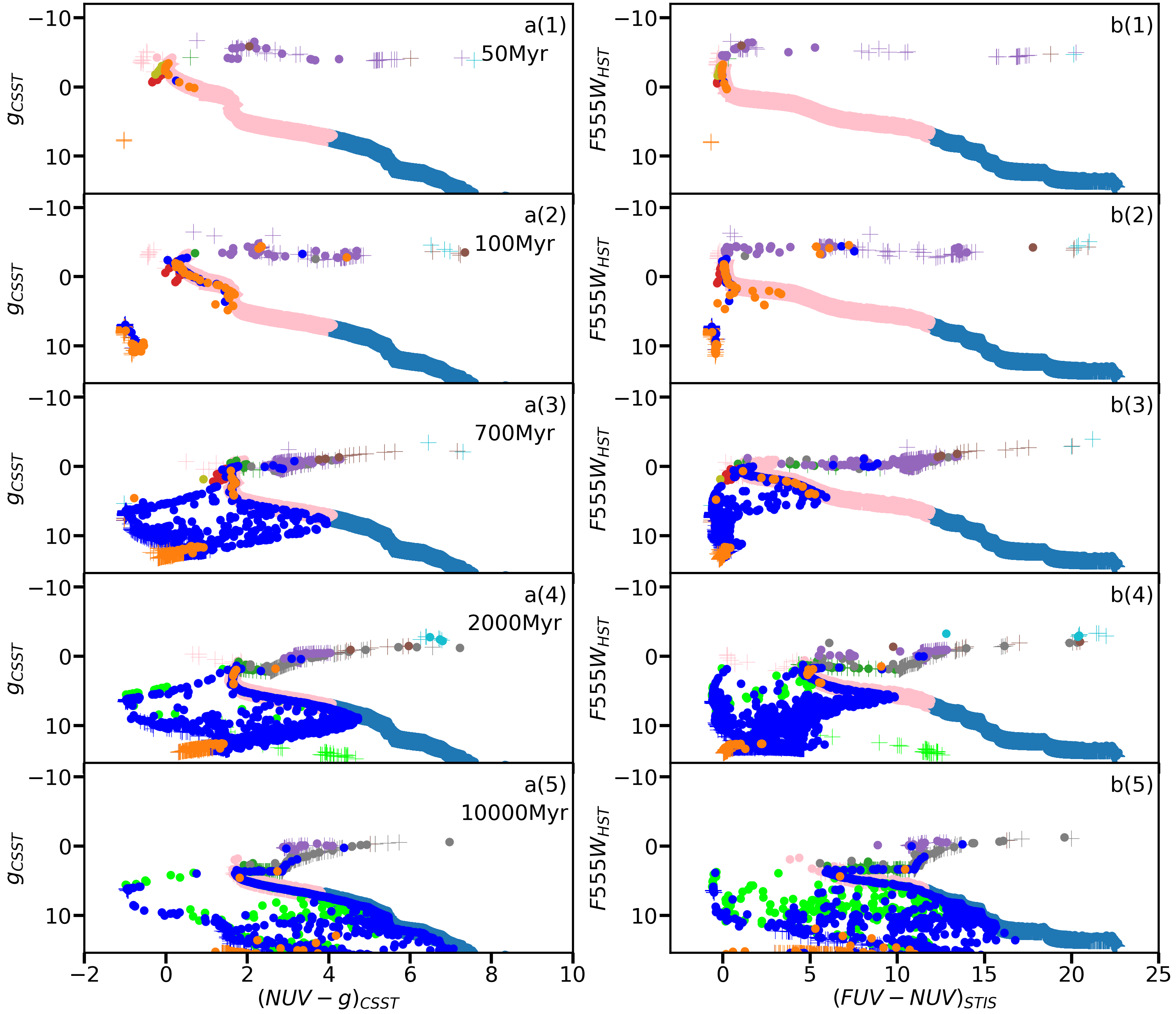}
    \caption{The CMD for model 100k-2Z. Symbols and colors are identical to those in Figure~\ref{fig:cmd_10k-2z}.}
\label{fig:cmd_100k-2z}
\end{figure*}

\begin{figure*}[tb!]
\centering
\includegraphics[angle=0, width=1.\textwidth]{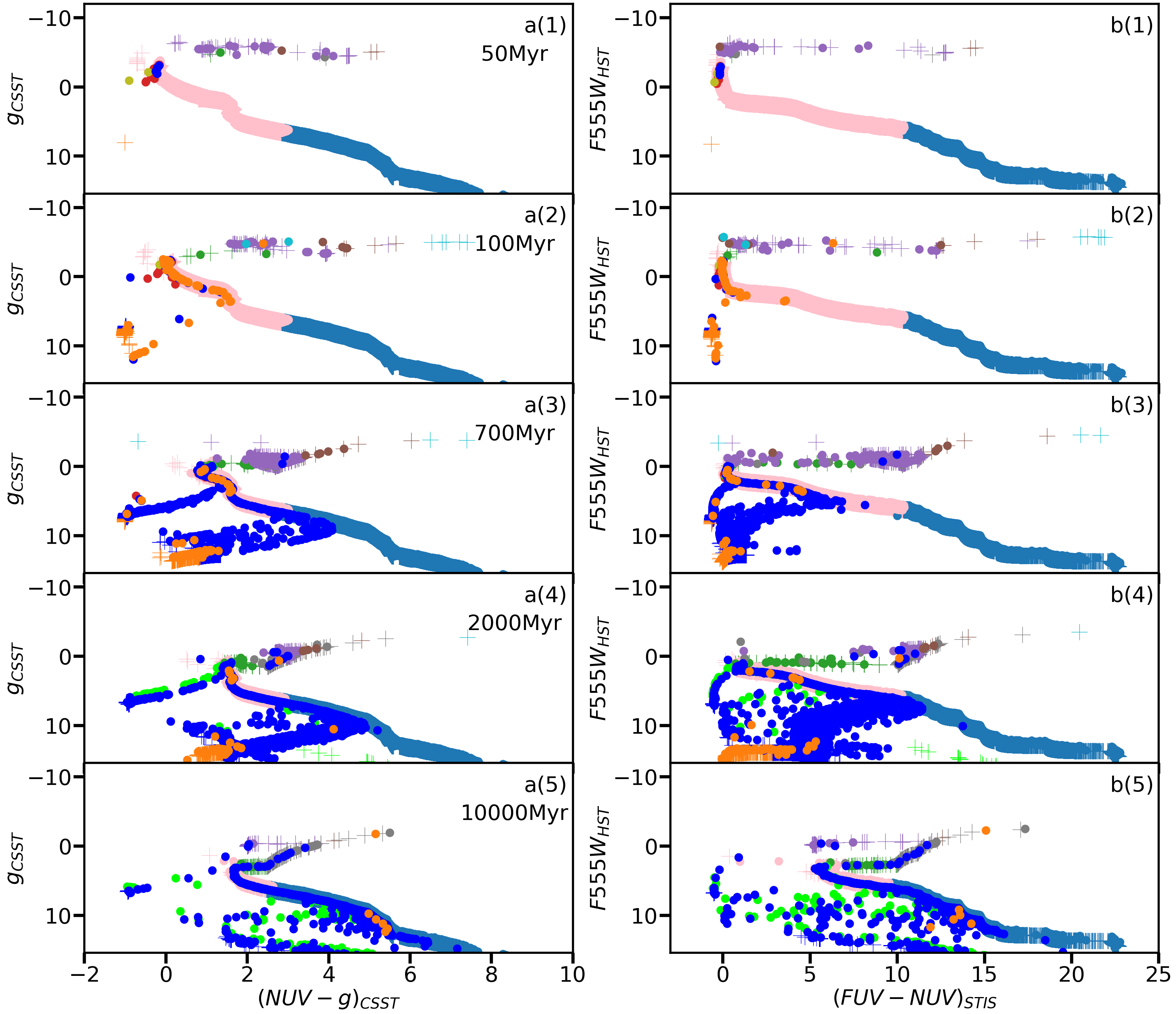}
    \caption{The CMDs for model 100k-3Z. Symbols and colors are identical to those in Figure~\ref{fig:cmd_10k-2z}.}
\label{fig:cmd_100k-3z}
\end{figure*}

\bibliographystyle{raa}
\bibliography{paper}

\end{document}